\documentclass[preprint,nofootinbib,tightenlines,superscriptaddress,amsfonts,amsmath,amssymb]{revtex4}
\usepackage{color}
\usepackage{bm}
\usepackage{graphicx}
\usepackage[hypertex]{hyperref}
\usepackage{multirow}%

\begin{document}
\baselineskip=15pt \parskip=4pt

\title{Exploring high-mass diphoton resonance without new colored states}
\author{Amine Ahriche}
\email{aahriche@ictp.it} \affiliation{Laboratory of Mathematical and
Sub-Atomic Physics (LPMPS), University of Constantine I, DZ-25000
Constantine, Algeria.}\affiliation{Department of Physics and Center
for Theoretical Sciences, National Taiwan University, Taipei 106,
Taiwan.}
\affiliation{The Abdus Salam International Centre for
Theoretical Physics, Strada Costiera 11, I-34014, Trieste, Italy.}
\author{Gaber Faisel}
\email{gfaisel@hep1.phys.ntu.edu.tw}\affiliation{Department of
Physics and Center for Theoretical Sciences, National Taiwan
University, Taipei 106, Taiwan.}\affiliation{Department of Physics,
Faculty of Arts and Sciences, S\"{u}leyman Demirel University,
Isparta, Turkey 32260.} \affiliation{Egyptian Center for Theoretical
Physics, Modern University for Information and Technology, Cairo
11212, Egypt.}
\author{Salah Nasri}
\email{snasri@uaeu.ac.ae} \affiliation{Department of Physics, United
Arab Emirates University, Al-Ain, UAE.}
\author{Jusak Tandean}
\email{jtandean@yahoo.com} \affiliation{Department of Physics and
Center for Theoretical Sciences, National Taiwan University, Taipei
106, Taiwan.} \affiliation{Physics Division, National Center for Theoretical
Sciences, Hsinchu 300, Taiwan\bigskip}

\begin{abstract}
A new heavy resonance may be observable at the LHC if it has a significant decay branching
fraction into a\,\,pair of photons.
We entertain this possibility by looking at the modest excess in the diphoton invariant
mass spectrum around 750 GeV recently reported in the ATLAS and CMS experiments.
Assuming that it is a spinless boson, dubbed $\tilde s$, we consider it within a model
containing two weak scalar doublets having zero vacuum expectation values and a scalar
singlet in addition to the doublet responsible for breaking the electroweak symmetry.
The model also possesses three Dirac neutral singlet fermions, the lightest one of which
can play the role of dark matter and which participate with the new doublet scalars in
generating light neutrino masses radiatively.
We show that the model is consistent with all phenomenological constraints and can
yield a production cross section $\sigma(pp\rightarrow\tilde{s}\rightarrow\gamma \gamma)$
of roughly the desired size, mainly via the photon-fusion contribution, without
involving extra colored fermions or bosons.
We also discuss other major decay modes of $\tilde s$ which are potentially testable in
upcoming LHC measurements.
\end{abstract}

\maketitle

\section{Introduction\label{intro}}

Some of the recent data collected at the LHC from proton-proton collisions at a
center-of-mass energy of \,$\sqrt s=13$\,TeV\, have turned up tantalizing
potential hints of physics beyond the standard model (SM). Specifically, upon
searching for new resonances decaying into two photons, the ATLAS and CMS
Collaborations~\cite{atlas:s2gg,cms:s2gg} have reported observing modest
excesses above the backgrounds peaked at a mass value of around 750\,\,GeV
with local (global) significances of 3.9$\sigma$ and\,\,3.4$\sigma$
(2.1$\sigma$ and\,\,1.6$\sigma$),
respectively~\cite{Aaboud:2016tru,Khachatryan:2016hje}. If interpreted as
telltales of a resonance, the ATLAS data suggest that it has a width of about
50\,GeV,\, whereas the CMS results prefer it to be
narrower~\cite{Aaboud:2016tru,Khachatryan:2016hje}. As pointed out in a
number of theoretical works~\cite{pp2s2gg-1} appearing very shortly after
the ATLAS and CMS announcements~\cite{atlas:s2gg,cms:s2gg}, the cross section
of producing the putative heavy particle decaying into $\gamma\gamma$ falls
within the range of roughly 2-13 fb, and it is possible for its width to be
less than 50\,GeV or even narrow.

Given the limited statistics of the diphoton excess events, it would still be
premature to hold a\,\,definite view concerning these findings. Nevertheless,
if the tentative indications of the existence of a non-SM state are confirmed
by upcoming measurements, the acquired data will not only constitute more
conclusive evidence for new physics, but also paint a clearer picture of the
new particle's properties which will then serve as a test for models. It is
therefore of interest in the meantime to explore a variety of new-physics
scenarios that can accommodate it, subject to the relevant available
experimental constraints, and also to look at other aspects of these recent
LHC results~\cite{pp2s2gg-1,pp2s2gg-2,Falkowski:2015swt,
Fichet:2015vvy,Csaki:2015vek,Csaki:2016raa,gammaF}.

Here we consider the possibility that the excess diphoton events proceeded
from the decay of a\,\,new spinless boson, which we denote by $\tilde{s}$ and
arises due to the presence of a complex scalar field,\,\,$\zeta$, transforming as
a singlet under the SM gauge group, SU(2)$_{L}\times$U(1)$_{Y}$. In our
scenario of interest, the scalar fields also include two new weak doublets,
$\eta_{1}^{}$ and $\eta_{2}^{}$, having zero vacuum expectation values
(VEVs), besides the doublet, $\Phi$, which contains the
Higgs boson in the SM. Moreover, the gauge sector is somewhat expanded in
comparison to that of the SM by the addition of a new Abelian gauge
symmetry,\thinspace\thinspace\textrm{U(1)}$_{D}$, under which $\zeta$ and
$\eta_{1,2}^{}$ are charged, while SM particles are not. Consequently,
$\eta_{1,2}^{}$ have no direct interactions with a pair of exclusively SM
fermions, whereas $\tilde{s}$ can couple at tree level to the latter because
of mixing between the remaining components of $\zeta$ and $\Phi$ after they
develop nonzero VEVs. Having no VEVs nor couplings to SM fermion pairs,
$\eta_{1,2}^{}$ have been termed inert in the literature~\cite{Keus:2013hya},
but being weak doublets they do interact directly with SM gauge bosons. For
simplification, we suppose that the \textrm{U(1)}$_{D}$ gauge boson has
vanishing kinetic mixing with the hypercharge gauge boson, and thus the former
can be regarded as dark.
We further assume that all these bosons belong to a more expanded
model that possesses three extra fermions $(N_1,N_2,N_3)$ which are Dirac in nature,
charged under\,\,U(1)$_{D}$, and singlet under the SM gauge group \cite{Ma:2013yga}.
The lightest mass eigenstate among the new fermions can serve as a\,\,dark matter (DM)
candidate if it is also lighter than the inert scalars, and both these fermions
and scalars participate in generating light neutrino masses at the loop level.
It is worth noting that in the absence of the singlets, $\zeta$ and $N_{1,2,3}$,
the model corresponds to one of the possible three-scalar-doublet cases cataloged in
Ref.\thinspace\thinspace\cite{Keus:2013hya} and has been examined for its interesting
potential impact on the Higgs trilinear coupling and electroweak phase transition
in\thinspace\thinspace Ref.\thinspace\thinspace\cite{Ahriche:2015mea}.

The remainder of the paper is organized as follows. In the next section, we
describe the salient features of the model and the nonstandard particles'
interactions of concern and masses. In\,\,Sec.\,\ref{sdecays}, we enumerate the
major decay modes of $\tilde s$. In Sec.\,\ref{constraints}, we discuss
constraints on the scalars from theoretical requirements, electroweak
precision data, and collider measurements.
In\,\,Sec.\,\ref{fdm}, we address the requirements on the lightest one of the new
fermions being the DM, how they in conjunction with the inert scalars can give rise to
loop-induced Majorana masses of the light neutrinos, and the implications for lepton flavor
violation and the muon anomalous magnetic moment.
We present our numerical analysis in Sec.\,\ref{numres}, demonstrating that the model can
generate the requisite LHC values of the production cross-section $\sigma(pp\to\tilde s\to
\gamma\gamma)$ mainly via the photon-fusion contribution. Hence our scenario
does not involve any colored fermions or bosons to enhance the $\tilde
s\gamma\gamma$ coupling. Also, we briefly discuss what other decay modes of
$\tilde s$ and additional signatures of the model may be checked
experimentally in order to probe the model more stringently. We give our
conclusions in Sec.\,\ref{conclusion}. Some complementary information and
formulas are relegated to a few appendices.

\section{Model\label{model}}

\begin{table}[b] \medskip
\begin{tabular}[c]{|c||c|c|c|c|c|c|c|c|c|}\hline
~~ & $\Phi$ & $\eta_1^{}$ & $\eta_2^{}$ & $\zeta$ & $L_k^{}$ & $N_k^{}\vphantom{|_{|_|}^|}$ \\
\hline\hline
SU(2)$_L\vphantom{|_|^|}$ &   2 &   2 &    2 & 1 & 2 & 1 \\ \hline
U(1)$_Y\vphantom{|_|^|}$  & 1/2 & 1/2 &  1/2 & 0 & $-$1/2 & 0 \\ \hline
\,U(1$)_D\vphantom{|_|^|}$ $[Z_2]$\, & \,0 $[+]$\, & \,1 $[-]$\, & \,$-$1 $[-]$\, & \,2 $[+]$\,
& \,0 [+]\, & \,1 $[-]$\, \\
\hline \end{tabular}
\caption{Charge assignments of the scalars, standard lepton doublets
\,$L_k=(\nu_k~~~\ell_k)^{\rm T}$,\, and new singlet fermions $N_k$ in the model,
for \,$k=1,2,3$.}%
\label{assignments}
\end{table}

The quantum numbers of the scalar, lepton doublet, and new Dirac singlet fermion fields are
listed in Table\thinspace\thinspace\ref{assignments}.
The gauge boson associated with \textrm{U(1)}$_{D}$ is referred to as $C$.
Accordingly, we can express the Lagrangian $\mathcal L$ describing their renormalizable
interactions with each other and with the SM gauge bosons, $W_{j}$ and $B$, as
\begin{align} \label{L}
\mathcal{L}\,= & \,(\mathcal{D}^{\rho}\Phi)^{\dagger}\mathcal{D}_{\rho}^{}\Phi
+ (\mathcal{D}^{\rho}\eta_{a}^{})^{\dagger}\mathcal{D}_{\rho}^{}\eta_{a}^{}
+ (\mathcal{D}^{\rho}\zeta)^{\dagger}\mathcal{D}_{\rho}^{}\zeta \,-\,\mathcal{V}
\nonumber \\ & +\,
\overline{N_k^{}}\big(i\slash\!\!\!\partial-g_D^{}\;\slash\!\!\!\!C\big)N_k^{}
\,-\, \tfrac{1}{2}_{\,\!}\varepsilon_{\,\!}B^{\rho\omega}C_{\rho\omega}
- \tfrac{1}{4} C^{\rho\omega}C_{\rho\omega} \,+\, {\cal L}_N^{}%
\,,\vphantom{|_{\int_|^|}^{}} \\
\mathcal{D}^{\rho}\Delta\,=  &  \, \big( \partial^{\rho}+\tfrac{i}{2}%
g\tau_{j}^{}W_{j}^{\rho}+ig_{Y}^{}\mathcal{Q}_{Y}^{}B^{\rho}+ig_{D}^{}\mathcal{Q}%
_{C\,}^{}C^{\rho} \big)  \Delta\,,~~~~~~~\Delta\,=\,\Phi,\eta_{1}^{},\eta
_{2}^{}\,, \nonumber \\
\mathcal{D}^{\rho}\zeta\,=  &  \, \big( \partial^{\rho}+ig_{D}^{}\mathcal{Q}%
_{C\,}^{}C^{\rho} \big)  \zeta\,,~~~~~~~\mathcal{Q}_{C}^{}(\Phi,\eta_{1}^{},\eta
_{2}^{},\zeta)\,=\,(0,\eta_{1}^{},-\eta_{2}^{},2\zeta
)\,, \vphantom{|_{\int_|^|}^{}} \nonumber \\ \label{LN}
{\cal L}_N^{} \,=  & ~\mbox{$-(M_N)_{kl}^{}$}\,\overline{N_k}\,P_R^{}N_l^{} \,-\,
({\cal Y}_1^{})_{kl}^{}\,\overline{L_k}\,\tilde\eta_1^{}P_R^{}N_l^{} -
({\cal Y}_2^{})_{kl}^{}\,\overline{L_k}\,\tilde\eta_2^{}P_R^{}N_l^{\rm c} \nonumber \\
& -\big(\hat{\texttt Y}_1^{}\big)_{kl}\,\overline{N_k^{\rm c}}\,P_R^{}N_l^{}\,\zeta^\dagger -
\big(\hat{\texttt Y}_2^{}\big)_{kl}\,\overline{N_k^{}}\,P_R^{}N_l^{\rm c\,}\zeta \;+\; {\rm H.c.} \,,
\vphantom{|_{\int_|^|}^{}} \\ \label{potential}
{\mathcal{V}}\,=  &  ~\mu_{1\,}^{2}\Phi^{\dagger}\Phi+\mu_{2a\,}^{2}\eta
_{a}^{\dagger}\eta_{a}^{}+\mu_{\zeta\,}^{2}|\zeta|^{2}+\tfrac{1}{2}%
\lambda_{1}(\Phi^{\dagger}\Phi)^{2}+\tfrac{1}{2}\lambda_{2a}(\eta_{a}%
^{\dagger}\eta_{a}^{})^{2}+\tfrac{1}{2}\lambda_{\zeta}|\zeta|^{4}\nonumber\\
&  +\lambda_{3a\,}\Phi^{\dagger}\Phi\,\eta_{a}^{\dagger}\eta_{a}^{}%
+\lambda_{3\zeta\,}\Phi^{\dagger}\Phi\,|\zeta|^{2}+\lambda_{4a\,}\Phi
^{\dagger}\eta_{a\,}^{}\eta_{a}^{\dagger}\Phi+\tfrac{1}{2}\left(
\lambda_{5\,}\Phi^{\dagger}\eta_{1\,}^{}\Phi^{\dagger}\eta_{2}^{{}%
}+\mathrm{H.c.}\right)  ~~~~~\nonumber\\
&  +\lambda_{6\,}\eta_{1}^{\dagger}\eta_{1\,}^{}\eta_{2}^{\dagger}\eta
_{2}^{}+\lambda_{7\,}\eta_{1}^{\dagger}\eta_{2\,}^{}\eta_{2}^{\dagger}%
\eta_{1}^{}+\lambda_{a\zeta\,}^{}\eta_{a}^{\dagger}\eta_{a\,}^{}%
|\zeta|^{2}+\left(  \mu_{\eta\zeta\,}^{}\eta_{1}^{\dagger}\eta_{2\,}^{{}%
}\zeta+\mathrm{H.c.}\right)  ,
\end{align}
where $g_{D}^{}$ and $\mathcal{Q}_{C}$ are the coupling constant and charge operator
of \textrm{U(1)}$_{D}$, respectively, $\varepsilon$ parameterizes the tree-level kinetic
mixing between the \textrm{U(1)}$_{Y,D}$ gauge bosons, $M_N$ is the Dirac mass matrix of
the singlet fermions, ${\cal Y}_{1,2}$ and $\hat{\texttt Y}_{1,2}$ are Yukawa coupling
matrices, \,$\tilde\eta_a^{}=i\tau_2^{}\eta_a^*$,\, summation over \,$a=1,2$\, and
\,$j,k,l=1,2,3$\, is implicit, \,$P_R=\tfrac{1}{2}(1+\gamma_5^{})$,\, and, after
electroweak symmetry breaking, in the unitary gauge
\begin{equation}
\Phi\,=\left(
\begin{array}
[c]{c}%
0\\
\frac{1}{\sqrt{2}}\left(  v+\phi\right)
\end{array}
\right)  ,~~~~~~~\eta_{a}\,=\left(
\begin{array}
[c]{c}%
\eta_{a}^{+}\vspace{3pt}\\
\eta_{a}^{0}%
\end{array}
\right)  ,~~~~\eta_{a}^{0}\,=\,\frac{\mathrm{Re}\,\eta_{a}^{0}+i_{\,}%
\mathrm{Im}\,\eta_{a}^{0}}{\sqrt{2}}\,,~~~~~~~\zeta\,=\,\frac{\tilde
{v}+\varsigma}{\sqrt{2}}\,,
\end{equation}
with $v$ and $\tilde{v}$ denoting the vacuum expectation values (VEVs) of
$\Phi$ and $\zeta$, respectively. The Hermiticity of $\mathcal{V}$ implies
that $\mu_{1,2a,\zeta}^{2}$ and $\lambda_{1,2a,3a,4a,6,7,\zeta,3\zeta,a\zeta}
$ must be real. Since the phases of $\eta_{1,2}^{}$ relative to $\Phi$ and
$\zeta$ can be arranged to render $\lambda_{5}$ and $\mu_{\eta\zeta}$ real,
without loss of generality we will choose these parameters to be real.
We can also pick a convenient basis such that $M_N$ is diagonal,
\,$M_N=(M_1,M_2,M_3)$.\,

One can see from Eq.\thinspace(\ref{potential}) that, after $\Phi$ and $\zeta$
develop nonzero VEVs, their remaining components $\phi$ and $\varsigma$,
respectively, generally mix with each other. Moreover, upon the VEV of $\zeta$ being
nonzero, the $\hat{\texttt Y}_{1,2}$ and $\mu_{\eta\zeta}$ terms break \textrm{U(1)}$_{D}$
into $Z_{2}$ under which the new fermions and inert scalars are odd,
as Table\thinspace\thinspace\ref{assignments} indicates, and all the other fields even.
Although the lightest electrically neutral $Z_{2}$-odd scalar is stable if it is also
lighter than $N_{1,2,3}$, we find that in our parameter space of interest it cannot be
a good DM candidate. This is because its annihilation into SM particles is too fast due
to its tree-level interactions with SM gauge and Higgs bosons and hence cannot produce
enough relic abundance.
On the other hand, if the lightest mass eigenstate among the new fermions is
also lighter than the inert scalars, it can play the role of DM, as we will discuss in more
detail later.
In the rest of this section and the following two sections we focus on the new scalars'
interactions and masses, while in Sec.\,\ref{fdm} we look at important implications of
the new fermions' presence.

After the \textrm{U(1)}$_{D}\rightarrow Z_{2}$ breaking, the
$\mu_{\eta\zeta}$ terms also induce the mixing of $Z_{2}$-odd
scalars of the same electric charge. To examine this more closely, we can
write the part of $\mathcal{L}$ from $\mathcal{V}$ which is quadratic in the
scalar fields as
\begin{equation}
\mathcal{L}\,\supset\,-\tfrac{1}{2}\left(  \phi~~~\varsigma\right)
\,M_{\phi\varsigma}^{2}\left(
\begin{array}
[c]{c}%
\phi\\
\varsigma
\end{array}
\right)  -\,\left(  \eta_{1}^{-}~~~\eta_{2}^{-}\right)  \,M_{\mathcal{C}}%
^{2}\left(
\begin{array}
[c]{c}%
\eta_{1}^{+} \vspace{3pt} \\
\eta_{2}^{+}%
\end{array}
\right)  -\,\tfrac{1}{2}\,\eta_{0}^{\mathrm{T}}M_{0}^{2}\,\eta_{0}\,,
\label{L2}%
\end{equation}
where the expressions for the matrices $M_{\phi\varsigma}^{2}$,
$M_{\mathcal{C}}^{2}$, $M_{0}^{2}$, and $\eta_{0}$ can be found in
Appendix\thinspace\thinspace\ref{app}.

Upon diagonalizing $M_{\phi\varsigma}^{2}$, we obtain the mass eigenstates $h
$ and $\tilde{s}$ and their respective masses $m_{h}$ and $m_{\tilde{s}}$
given by
\begin{align}
\left(  \!%
\begin{array}
[c]{c}%
\phi\\
\varsigma
\end{array}
\!\right)   &  =\left(  \!%
\begin{array}
[c]{ccc}%
c_{\xi} &  & s_{\xi}\\
-s_{\xi} &  & c_{\xi}%
\end{array}
\!\right)  \left(  \!%
\begin{array}
[c]{c}%
h\\
\tilde{s}%
\end{array}
\!\right)  \equiv\,\mathcal{O}_{\phi\varsigma}\left(  \!%
\begin{array}
[c]{c}%
h\\
\tilde{s}%
\end{array}
\!\right)  ,~~~~~~~c_{\xi}\,=\,\cos\xi\,,~~~s_{\xi}\,=\,\sin\xi\,,\nonumber\\
\vphantom{|_{\int}^{\int_|}} &  \mathcal{O}_{\phi\varsigma
\,}^{\mathrm{T}}M_{\phi\varsigma\,}^{2}\mathcal{O}_{\phi\varsigma}%
^{}\,=\,\mathrm{diag}\left(  m_{h}^{2},m_{\tilde{s}}^{2}\right)
,\nonumber\label{m2hs}\\
2m_{h,\tilde{s}}^{2} \,  &  =\, m_{\phi}^{2}+m_{\varsigma}^{2}\mp\sqrt{\left(
m_{\phi}^{2}-m_{\varsigma}^{2}\right)  ^{2}+m_{\phi\varsigma}^{4}%
}\,,~~~~~~~\tan(2\xi)\,=\,\frac{m_{\phi\varsigma}^{2}}{m_{\varsigma}%
^{2}-m_{\phi}^{2}}\,,\nonumber\\
m_{\phi}^{2} \,  &  =\, \lambda_{1} v^{2} \,, ~~~~~ m_{\varsigma}^{2} \,=\,
\lambda_{\zeta} \tilde{v}^{2} \,, ~~~~~ m_{\phi\varsigma}^{2} \,=\,
2\lambda_{3\zeta}^{~~\;}v\tilde{v} \,.
\end{align}
It follows that \,$m_{h}\sim125$\,GeV\, and \,$m_{\tilde s}\sim750$\,GeV.\,
Furthermore, all the tree-level couplings of $h$ ($\tilde{s}$) to SM fermions
and weak bosons, $W$ and $Z$, are $c_{\xi}$ $\left(  s_{\xi}\right)  $ times
the corresponding SM Higgs couplings.

For the electrically charged inert scalars, from the $M_{\mathcal{C}}^{2}$
term in Eq.\thinspace(\ref{L2}), we arrive at the mass eigenstates
$H_{1,2}^{\pm}$ and their masses $m_{H_{1},H_{2}}$ given by
\begin{align}
\left(\!\begin{array}[c]{c} \eta_{1}^{+} \vspace{5pt} \\ \eta_{2}^{+} \end{array}\!\right)
& = \left(\!\begin{array}[c]{ccc} c_H^{} && s_H^{} \vspace{2pt} \\
-s_H^{} && c_H^{} \end{array}\!\right)  \left(\!\begin{array}[c]{c} H_1^+ \vspace{5pt} \\
H_{2}^{+} \end{array}\!\right)  \equiv\, \mathcal{U}_{\mathcal C} \left(\!\begin{array}[c]{c}
H_{1}^{+} \vspace{5pt} \\ H_{2}^{+} \end{array}\!\right) , ~~~~~~~
c_H^{} \,=\, \cos\theta_H^{} \,, ~~~ s_H^{} \,=\, \sin\theta_H^{} \,,
\nonumber \label{m2H} \\
\vphantom{|_{\int_|}^{\int_|^|}} & \mathcal{U}_{\mathcal{C}\,}^\dagger M_{\mathcal C}^2\,
\mathcal{U}_{\mathcal C}^{} \,=\, \mathrm{diag} \left( m_{H_1}^2 ,_{\,\!} m_{H_2}^2 \right) ,
~~~~~~~ m_{H_a}^{} \equiv\, m_{H_{a}^{\pm}} \,,
\nonumber \\
2m_{H_{1},H_{2}}^{2}\,  &  =\,m_{c_{1}}^{2}+m_{c_{2}}^{2}\mp\sqrt
{\big(m_{c_{1}}^{2}-m_{c_{2}}^{2}\big)^{2}+m_{c\zeta}^{4}}\,,~~~~~~~\tan
(2\theta_{H})\,=\,\frac{m_{c\zeta}^{2}}{m_{c_{2}}^{2}-m_{c_{1}}^{2}}\,,
\end{align}
where $m_{c_{a},c\zeta}^{2}$ are related to other parameters in Eq.\thinspace
(\ref{M0}) and we have taken $\mu_{\eta\zeta}^{}$, and hence $m_{c\zeta}%
^{2}$, to be real. Similarly, the mixing of the electrically neutral inert
scalars gives rise to the mass eigenstates $\mathcal{S}_{a}$ and
$\mathcal{P}_{a}$ with their respective masses $m_{\mathcal{S}_{a}}$ and
$m_{\mathcal{P}_{a}}$ according to
\begin{align}
\left(\begin{array}[c]{c} \mathrm{Re}\,\eta_{1}^{0} \vspace{3pt} \\
\mathrm{Re}\,\eta_{2}^{0}\vspace{3pt} \\ \mathrm{Im}\,\eta_{1}^{0} \vspace{3pt} \\
\mathrm{Im}\,\eta_{2}^{0} \end{array}\right)
&  =\left(\begin{array}[c]{ccccccc}
 c_S^{} && s_S^{} && 0 && 0 \vspace{1pt} \\ -s_S^{} && c_S^{} && 0 && 0 \vspace{1pt} \\
0 && 0 && c_P^{} &&  s_P^{} \vspace{1pt} \\ 0 && 0 && s_P^{} && -c_P^{} \end{array}\right)
\left(\begin{array}[c]{c}
\mathcal{S}_1^{} \vspace{2pt} \\ \mathcal{S}_2^{} \vspace{2pt} \\
\mathcal{P}_1^{} \vspace{2pt} \\ \mathcal{P}_2^{} \end{array}\right)
\equiv\, {\mathcal O}_0^{} \left(\begin{array}[c]{c} \mathcal{S}_1^{} \vspace{2pt} \\
\mathcal{S}_2^{} \vspace{2pt} \\ \mathcal{P}_1^{} \vspace{2pt} \\
\mathcal{P}_2^{} \end{array}\right)  ,
\nonumber \label{mSmP} \\ \vphantom{|^{\int}}
c_S^{}  & =\, \cos\theta_S^{} \,, ~~~ s_S^{} \,=\, \sin\theta_S^{}\,, ~~~~~~~
c_P^{} \,=\, \cos\theta_P^{} \,, ~~~ s_P^{} \,=\, \sin\theta_P^{} \,,
\nonumber \\ \vphantom{|_{\int^|}^{\int}}
& \mathcal{O}_{0\,}^{\mathrm T} M_{0\,}^2\mathcal{O}_{0}^{}
\,=\, \mathrm{diag} \left( m_{\mathcal{S}_1}^2,_{\,\!}m_{\mathcal{S}_2}^2,_{\,\!}
m_{\mathcal{P}_1}^{2},_{\,\!}m_{\mathcal{P}_2}^2\right) ,
\nonumber \\
2m_{\mathcal{S}_1,\mathcal{S}_2}^2 & =\, m_{n_1}^2+m_{n_2}^2 \mp
\sqrt{\left( m_{n_{1}}^{2}-m_{n_{2}}^2 \right)^2+m_{n\zeta}^4} \,, ~~~~~
\tan(2\theta_{S})\,=\,\frac{m_{n\zeta}^{2}}{m_{n_{2}}^{2}-m_{n_{1}}^{2}\vphantom{|_{\int}^{}}} \,,
\nonumber \\
2m_{\mathcal{P}_{1},\mathcal{P}_{2}}^{2}  &  =m_{n_{1}}^{2}+m_{n_{2}}^{2}%
\mp\sqrt{\left(  m_{n_{1}}^{2}-m_{n_{2}}^{2}\right)  ^{2}+\tilde{m}_{n\zeta
}^{4}}\,,~~~~~\tan(2\theta_{P})\,=\,\frac{\tilde{m}_{n\zeta}^{2}}{m_{n_{2}%
}^{2}-m_{n_{1}}^{2}}\,,
\end{align}
where $m_{n_{a}}^{2}$, $m_{n\zeta}^{2}$, and $\tilde{m}_{n\zeta}^{2}$ are
defined in Eq.\thinspace(\ref{m2nz}). 
From the last two lines we get
\[
m_{\mathcal{S}_{1}}^{2}+m_{\mathcal{S}_{2}}^{2}\;=\;m_{\mathcal{P}_{1}}%
^{2}+m_{\mathcal{P}_{2}}^{2}\,.
\]
The simple form of $\mathcal{O}_{0}$ above is due to $\mu_{\eta\zeta}^{}$
again as well as $\lambda_{5}$, and hence $m_{n\zeta}^{2}$ and $\tilde
{m}_{n\zeta}^{2}$, being real, which in view of Eq.\thinspace(\ref{M0}) also
implies that
\[
m_{n\zeta}^{2}\,=\,\tilde{m}_{n\zeta}^{2}+2\,m_{c\zeta}^{2}\,.
\]

The kinetic portion of $\mathcal{L}$ in Eq.\thinspace(\ref{L})
contains the interactions of the scalars with the SM gauge bosons,
\begin{align}
{\mathcal{L}}\,\supset\,  &  ~\frac{g}{2}\!%
\begin{array}
[t]{l}%
\big\{ \big[ (c_{H}c_{S}+s_{H}s_{S})\, i \big(H_{1}^{+}%
\mbox{\footnotesize$\stackrel{\scriptscriptstyle\leftrightarrow}{\partial}$}{}%
^{\mu}{\mathcal{S}}_{1}^{}+H_{2}^{+}%
\mbox{\footnotesize$\stackrel{\scriptscriptstyle\leftrightarrow}{\partial}$}{}%
^{\mu}{\mathcal{S}}_{2}^{} \big)
+ (c_{H}s_{S}-s_{H}c_{S})\, i \big( H_{1}^{+}%
\mbox{\footnotesize$\stackrel{\scriptscriptstyle\leftrightarrow}{\partial}$}{}%
^{\mu}{\mathcal{S}}_{2}^{} - H_{2}^{+}%
\mbox{\footnotesize$\stackrel{\scriptscriptstyle\leftrightarrow}{\partial}$}{}%
^{\mu}{\mathcal{S}}_{1}^{}\big) \vspace{2pt}\\
~~ + (c_{H}c_{P}-s_{H}s_{P}) \big( H_{1}^{+}%
\mbox{\footnotesize$\stackrel{\scriptscriptstyle\leftrightarrow}{\partial}$}{}%
^{\mu}{\mathcal{P}}_{1}-H_{2}^{+}%
\mbox{\footnotesize$\stackrel{\scriptscriptstyle\leftrightarrow}{\partial}$}{}%
^{\mu}{\mathcal{P}}_{2}^{}\big)
+ (c_{H}s_{P}+s_{H}c_{P}) \big(  H_{1}^{+}%
\mbox{\footnotesize$\stackrel{\scriptscriptstyle\leftrightarrow}{\partial}$}{}%
^{\mu}{\mathcal{P}}_{2}^{}+H_{2}^{+}%
\mbox{\footnotesize$\stackrel{\scriptscriptstyle\leftrightarrow}{\partial}$}{}%
^{\mu}{\mathcal{P}}_{1}^{}\big)\big]W_{\mu}^{-} \vspace{2pt} \\
~+\, {\mathrm{H.c.}} \big\}
\end{array}
\nonumber\label{Lgauge}\\
&  \!\!\!\! +\, \frac{g}{2c_{\mathrm w}^{}}\!\begin{array}[t]{l}%
\left[  \left(  c_{S}c_{P}-s_{S}s_{P}\right)  \left(  {\mathcal{P}}_{1}%
\mbox{\footnotesize$\stackrel{\scriptscriptstyle\leftrightarrow}{\partial}$}{}%
^{\mu}{\mathcal{S}}_{1}-{\mathcal{P}}_{2}%
\mbox{\footnotesize$\stackrel{\scriptscriptstyle\leftrightarrow}{\partial}$}{}%
^{\mu}{\mathcal{S}}_{2}\right)  \right.  
+(c_{S}s_{P}+s_{S}c_{P})\left(  {\mathcal{P}}_{1}%
\mbox{\footnotesize$\stackrel{\scriptscriptstyle\leftrightarrow}{\partial}$}{}%
^{\mu}{\mathcal{S}}_{2}+{\mathcal{P}}_{2}%
\mbox{\footnotesize$\stackrel{\scriptscriptstyle\leftrightarrow}{\partial}$}{}%
^{\mu}{\mathcal{S}}_{1}\big)\right]  Z_{\mu}%
\end{array}
\nonumber\\
&  \!\!\!\! +\, i\big( H_{1}^{+}%
\mbox{\footnotesize$\stackrel{\scriptscriptstyle\leftrightarrow}{\partial}$}{}%
^{\mu}H_{1}^{-}+H_{2}^{+}%
\mbox{\footnotesize$\stackrel{\scriptscriptstyle\leftrightarrow}{\partial}$}{}%
^{\mu}H_{2}^{-}\bigr) \bigl(  eA_{\mu}-g_{L}^{}Z_{\mu}\bigr)
+ \big( H_{1}^{+}H_{1}^{-}+H_{2}^{+}H_{2}^{-}\big) \bigg[
\frac{g^{2}}{2}W^{+\mu}W_{\mu}^{-}+\big( eA-g_{L}^{}Z\big) ^{\!2}\bigg]
\nonumber\\
&  \!\!\!\! +\, \frac{g^{2}}{4}\left[  \left(  c_{\xi}h+s_{\xi}\tilde{s}+v\right)
^{2}+{\mathcal{S}}_{1}^{2}+{\mathcal{S}}_{2}^{2}+{\mathcal{P}}_{1}%
^{2}+{\mathcal{P}}_{2}^{2}\right]  \left(  W^{+\mu}W_{\mu}^{-}+\frac{Z^{2}%
}{2c_{\mathrm{w}}^{2}}\right)  ,
\end{align}
where
\begin{equation}
X\raisebox{1pt}{\footnotesize$\stackrel{\scriptscriptstyle\leftrightarrow}{\partial}$}{}%
^{\mu}Y\,=\,X\partial^{\mu}Y-Y\partial^{\mu}X\,,~~~~~g_{L}^{}\,=\,\frac
{g}{2c_{\mathrm w}^{}}\left(  2s_{\mathrm{w}}^{2}-1\right)  ,~~~~~
c_{\mathrm w}^{} \,=\, \cos\theta_{\mathrm w}^{}\,=\,\sqrt{1-s_{\mathrm w}^{2}}\;,
\end{equation}
with $\theta_{\mathrm{w}}$ being the usual Weinberg angle. These affect the
oblique electroweak parameters, to be treated later on.

From Eq.\,(\ref{Lgauge}), one can see that at tree level the masses of the $W
$ and $Z$ bosons are related to $v$ by \,$m_W^{} =c_{\mathrm w}^{}m_Z^{} =g
v/2$,\, just as in the SM. Although not displayed, there are also terms for
the interactions of $\eta_{a}^{}$ with the dark gauge boson $C$, from which we
obtain its mass to be~\,$m_{C}^{}=2 g_{D}^{}\tilde{v}$.\,
Numerically, we assume that \,$m_{C}^{}>m_{\tilde s}^{}$,\, which is reasonable because
the preferred value of $\tilde v$ is at least a few TeV, as will be seen later.

The kinetic part of $\mathcal L$ in Eq.\,(\ref{L}) also contains the tree-level mixing
between $B$ and $C$ parameterized by $\varepsilon$, which can be of $\mathcal O$(1).
Since $\eta_{1,2}^{}$ carry both U(1)$_Y$ and U(1)$_D$ charges, these scalars give rise
to loop-induced kinetic mixing between $B$ and $C$.
For simplicity, we suppose that the sum of these tree- and loop-level contributions is such
that the kinetic mixing between $B$ and $C$ is negligible, as stated in Sec.\,\ref{intro}.

Now, from the potential in Eq.\thinspace(\ref{potential}), we derive
\begin{eqnarray}
\label{Llambdas}{\mathcal{L}} &\supset&
-\Bigl[ \tfrac{1}{2}\lambda_{\tilde{s}hh}^{}h^{2}%
+\lambda_{\tilde{s}H_{a}H_{a}}^{}H_{a}^{+}H_{a}^{-}+\lambda_{\tilde{s}%
H_{1}H_{2}}^{}\left(  H_{1}^{+}H_{2}^{-}+H_{2}^{+}H_{1}^{-}\right)
\Bigr] \tilde s\tilde v\nonumber\\
&& \!-\; \Big(\tfrac{1}{2}\lambda_{\tilde{s}{\mathcal{S}}_{a}{\mathcal{S}}_{a}%
}{\mathcal{S}}_{a}^{2}+\tfrac{1}{2}\lambda_{\tilde{s}{\mathcal{P}}%
_{a}{\mathcal{P}}_{a}}{\mathcal{P}}_{a}^{2}+\lambda_{\tilde{s}{\mathcal{P}%
}_{1}{\mathcal{P}}_{2}}{\mathcal{P}}_{1}{\mathcal{P}}_{2}+\lambda_{\tilde
{s}{\mathcal{S}}_{1}{\mathcal{S}}_{2}}{\mathcal{S}}_{1}{\mathcal{S}}%
_{2}\Big) \tilde s\tilde v \nonumber\\
&& \!-\; \Big( \tfrac{1}{6} \lambda_{hhh}^{} h^2 + \lambda_{hH_{a}H_{a}}^{}H_{a}^{+}H_{a}^{-}
\Big) h v \,,
\end{eqnarray}
where summation over \thinspace$a=1,2 $\thinspace\ is implicit and the
formulas for the $\lambda$'s are given in Appendix\thinspace\thinspace\ref{app}.
These couplings determine the amplitudes for $\tilde{s}$ decays
into $hh$ or a pair of the inert scalars if kinematically allowed and, along
with Eq.\,(\ref{Lgauge}), are pertinent to $h$ and $\tilde{s}$ decays into
$\gamma\gamma$ and $\gamma Z$. These are some of the prominent decay channels
of $\tilde s$, to which we turn next.

\section{Decay modes of \boldmath$\tilde s$\label{sdecays}}

To examine the most important decay modes of $\tilde s$, we set its mass to be
\,$m_{\tilde s}=750$\,GeV\, for definiteness, whereas in the case of $h$ we assign
\,$m_{h}^{}=125.1$\,GeV,\, in accord with the latest mass determination\,\,\cite{lhc:mh}.
Hence $\tilde s$ can decay directly into $hh$ and, if kinematically permitted, into a pair of
the inert scalars or new singlet fermions.
With $\mathcal{X}$ and $\mathcal{Y}$ representing the two scalars in the final
state, for \,$m_{\tilde s}>m_{\mathcal{X}}+m_{\mathcal{Y}}$\, the decay rate is
\begin{equation}
\Gamma(\tilde{s}\rightarrow\mathcal{XY})\,=\,\frac{|\lambda_{\tilde
{s}{\mathcal{XY}}\,}\tilde{v}|^{2}}{(1+\delta_{\mathcal{XY}})16\pi
\,m_{\tilde{s}}^{3}}\sqrt{\big(m_{\tilde{s}}^{2}-m_{\mathcal{X}}%
^{2}-m_{\mathcal{Y}}^{2}\big)^{2}-4m_{\mathcal{X}}^{2}m_{\mathcal{Y}}^{2}}\,,
\end{equation}
where the $\lambda_{\tilde{s}\mathcal{XY}}$ expressions for various
$\mathcal{XY}$ pairs are collected in Eqs.\thinspace\thinspace(\ref{lhhh}%
)-(\ref{sPP}) and \,$\delta_{\mathcal{XY}}=1$\,(0)\, if \,$\mathcal{X}%
=\mathcal{Y}$ $(\mathcal{X}\neq\mathcal{Y})$.\, Thus, for instance,
\thinspace$\Gamma(\tilde{s}\rightarrow hh)\simeq1.25\times10^{-5}%
\,|\lambda_{\tilde{s}hh}\tilde{v}|^{2}$/\textrm{GeV}.\thinspace\ The $\tilde
s$ decays into final states containing 3 scalars may also happen, but such
channels have relatively much smaller rates due to phase-space suppression and
therefore can be neglected.
For the $\tilde s$ decay into the singlet fermions, the rate turns out to be small in
the parameter space of interest, and so we will neglect the effect of this channel on
the total width of $\tilde s$ hereafter.

Because of the $\phi$-$\varsigma$ mixing as specified in Eq.\,(\ref{m2hs}),
all the tree-level couplings of $h$ ($\tilde{s}$) to SM fermions, $W$, and $Z
$ are \,$c_{\xi}$ $\left(  s_{\xi}\right)  $ times the corresponding SM Higgs
couplings. It follows that, since a SM Higgs boson of mass 750\,GeV decays
almost entirely into \,$W^{+}W^{-},ZZ,t\bar{t}$\thinspace\ at rates which obey
the ratio \,$\Gamma\big(h_{\mathrm{SM}}^{750}\to W^{+}W^{-}\big):
\Gamma\big(h_{\mathrm{SM}}^{750}\to ZZ\big): \Gamma\big(h_{\mathrm{SM}}%
^{750}\to t\bar t\big)=145:72:30$\, and amount to \thinspace$247$%
\thinspace\textrm{GeV}~\cite{lhctwiki}, the rates of \,$\tilde s\to W^{+}%
W^{-},ZZ,t\bar{t}$\, conform to the same ratio,
\begin{equation}
\label{rateratio}\Gamma(\tilde{s}\rightarrow W^{+}W^{-}):\Gamma(\tilde
{s}\rightarrow ZZ): \Gamma(\tilde{s}\rightarrow t\bar{t}) \,=\, 145:72:30 \,,
\end{equation}
and sum up to
\begin{equation}
\Gamma(\tilde{s}\rightarrow W^{+}W^{-})+\Gamma(\tilde{s}\rightarrow ZZ)
+\Gamma(\tilde{s}\rightarrow t\bar{t}) \,=\, 247\;\mathrm{GeV}\,s_{\xi}^{2}\,.
\end{equation}

Our main channel of interest, \thinspace$\tilde s\rightarrow\gamma\gamma
$,\thinspace\ as well as \thinspace$\tilde s\rightarrow\gamma Z$%
,\thinspace\ arise from $t$, $W$, and $H_{1,2}$ loop diagrams, in analogy to
\thinspace$h\rightarrow\gamma\gamma$\thinspace\ and \thinspace$h\rightarrow
\gamma Z$,\thinspace\ respectively. In the absence of the singlet scalar, we
have derived the rates of the latter decays in Ref.\,\cite{Ahriche:2015mea}.
Modifying the rate formulas in the presence of $\tilde s$, we now have
\begin{align}
\Gamma(h\rightarrow\gamma\gamma)  &  \,=\, \frac{\alpha^{2}G_{\mathrm{F}}%
m_{h}^{3}}{128\sqrt{2}\,\pi^{3}}\left\vert \frac{4c_{\xi}}{3}A_{1/2}%
^{\gamma\gamma}(\kappa_{t})+c_{\xi}A_{1}^{\gamma\gamma}(\kappa_{W}%
)+\,\sum\limits_{a=1}^{2}\frac{\lambda_{hH_{a}H_{a}}^{}v^{2}}{2m_{H_{a}}^{2}%
}A_{0}^{\gamma\gamma}(\kappa_{H_{a}}) \right\vert ^{2}\,,
\nonumber\label{h2gz}\\
\nonumber\\
\Gamma(h\rightarrow\gamma Z)  &  \,=\,\frac{\alpha G_{\mathrm{F}}^{2}m_{W}%
^{2}\left(  m_{h}^{2}-m_{Z}^{2}\right)  ^{3}}{64\pi^{4}m_{h}^{3}}\left\vert
\frac{6-16s_{\mathrm{w}}^{2}}{3c_{\mathrm{w}}}\,c_{\xi}A_{1/2}^{\gamma
Z}(\kappa_{t},{\mathtt{z}}_{t})+c_{\xi}c_{\mathrm{w}}A_{1}^{\gamma Z}%
(\kappa_{W},{\mathtt{z}}_{W})\right. \nonumber\\
&  \hspace{25ex}\left.  -\;\frac{1-2s_{\mathrm{w}}^{2}}{c_{\mathrm{w}}}%
\,\sum\limits_{a=1}^{2}\frac{\lambda_{hH_{a}H_{a}}^{}v^{2}}{2m_{H_{a}}^{2}%
}A_{0}^{\gamma Z}(\kappa_{H_{a}},{\mathtt{z}}_{H_{a}}) \right\vert ^{2}\,,
\end{align}
where \,$\alpha=1/128$\, and $G_{\mathrm{F}}$ are the usual fine-structure and
Fermi constants, respectively, the expressions for the form factors
$A_{0,1/2,1}^{\gamma\gamma,\gamma Z}$ are available from Ref.\thinspace
\cite{Chen:2013vi}, the $A_{0}^{\gamma\gamma,\gamma Z}$ terms originate from
the $H_{1,2}$ loop diagrams, \thinspace$\kappa_{\beta}=4m_{\beta}^{2}%
/m_{h}^{2}$,\thinspace\ and \thinspace${\mathtt{z}}_{\beta}=4m_{\beta}%
^{2}/m_{Z}^{2}$.\thinspace\ Accordingly, we can deduce the rates of
\thinspace$\tilde{s}\rightarrow\gamma\gamma,\gamma Z$\thinspace\ to be
\begin{align}
\Gamma(\tilde{s}\rightarrow\gamma\gamma)  &  \,=\,\frac{\alpha^{2}%
G_{\mathrm{F}}m_{\tilde{s}}^{3}}{128\sqrt{2}\,\pi^{3}}\left\vert \frac
{4s_{\xi}}{3}A_{1/2}^{\gamma\gamma}\left(  \tilde{\kappa}_{t}\right)  +s_{\xi
}A_{1}^{\gamma\gamma}\left(  \tilde{\kappa}_{W}\right)  +\,\sum\limits_{a=1}%
^{2}\frac{\lambda_{\tilde{s}H_{a}H_{a}}^{}v\tilde{v}}{2m_{H_{a}}^{2}}%
A_{0}^{\gamma\gamma}(\tilde{\kappa}_{H_{a}}) \right\vert ^{2}%
\,,\nonumber\label{s2gz}\\
\nonumber\\
\Gamma(\tilde{s}\rightarrow\gamma Z)  & \,=\, \frac{\alpha G_{\mathrm F}^{2}m_{W}^{2} \left(
m_{\tilde{s}}^{2}-m_{Z}^{2}\right)^{3}}{64\pi^{4}m_{\tilde{s}}^{3}}
\left\vert \frac{6-16s_{\mathrm{w}}^{2}}{3c_{\mathrm w}^{}}\,s_\xi^{}A_{1/2}^{\gamma Z}
\left( \tilde{\kappa}_{t}^{}, {\mathtt z}_t^{}\right)
+ s_\xi^{}c_{\mathrm w}^{}A_{1}^{\gamma Z}(\tilde{\kappa}_W^{},{\mathtt z}_W^{}) \right. \nonumber\\
&  \hspace{25ex}\left.  -\;\frac{1-2s_{\mathrm{w}}^{2}}{c_{\mathrm w}^{}}\,
\sum\limits_{a=1}^{2}\frac{\lambda_{\tilde{s}H_{a}H_{a}}^{}v\tilde{v}}{2m_{H_{a}}^{2}}
A_{0}^{\gamma Z}(\tilde{\kappa}_{H_{a}},{\mathtt{z}}_{H_{a}})
\right\vert ^{2}\,,
\end{align}
where \thinspace$\tilde{\kappa}_\beta^{}=4m_{\beta}^{2}/m_{\tilde{s}}^{2}%
$\thinspace\ and in this case we set \,$\alpha=1/125$.\,

The aforementioned $s$ decay channels are the relevant contributions to
$\Gamma_{\tilde{s}}$. It follows that we can write for the branching fraction
of \thinspace$\tilde{s}\rightarrow\gamma\gamma$\thinspace\
\begin{align}
\label{B(s->gg)} &  \mathcal{B}(\tilde{s}\rightarrow\gamma\gamma
)\,=\,\frac{\Gamma_{\tilde{s}\rightarrow\gamma\gamma}}{\Gamma_{\tilde{s}}%
}\,,\nonumber\\
\Gamma_{\tilde{s}}\,\simeq\,\,  &  \Gamma(\tilde{s}\rightarrow\gamma
\gamma)+\Gamma(\tilde{s}\rightarrow\gamma Z) +\Gamma(\tilde{s}\rightarrow hh)
+247\;\mbox{GeV}\,s_{\xi}^{2}%
+\raisebox{3pt}{\scriptsize$\displaystyle\sum_{\rm
inert}$}\Gamma(\tilde{s}\rightarrow\mathcal{XY}) \,,
\end{align}
where in the last term of the second line the sum includes only decay modes with the inert
scalar masses satisfying \,$m_{\mathcal X}^{}+m_{\mathcal Y}^{}<m_{\tilde s}^{}$.\,
As mentioned above, it is also possible for $\tilde s$ to decay into a\,\,pair of the new
singlet fermions if they are sufficiently light, but in this study we concentrate on
the parameter space where their couplings to $\tilde s$ are small enough to make such
decay channels negligible.

\section{Constraints on new scalars\label{constraints}}

\subsection{Theoretical constraints\label{theory}}

The parameters in the scalar potential need to meet a number of theoretical
requirements. The stability of the vacuum implies that $\mathcal{V}$ must be
bounded from below. This entails that%
\begin{equation}
\lambda_{1},\lambda_{\zeta},\lambda_{21},\lambda_{22}>0\,, ~~~~~ \left\vert
\begin{array}
[c]{ccccc}%
\lambda_{{1}} &  & \lambda_{{1}}^{\prime} &  & \lambda_{{2}}^{\prime}\\
\lambda_{{1}}^{\prime} &  & \lambda_{{21}} &  & \lambda_{{3}}^{\prime}\\
\lambda_{{2}}^{\prime} &  & \lambda_{{6}}^{0}+\lambda_{{7}}^{0} &  &
\lambda_{{22}}%
\end{array}
\right\vert >0\,, ~~~~~ \left\vert
\begin{array}
[c]{ccccccc}%
\lambda_{{1}} &  & \,\lambda_{3\zeta}^{0} &  & \,\lambda_{{1}}^{\prime} &  &
\lambda_{{2}}^{\prime}\\
\lambda_{3\zeta}^{0} &  & \,\lambda_{{x}} &  & \,\lambda_{1\zeta}^{0} &  &
\,\lambda_{2\zeta}^{0}\\
\,\lambda_{{1}}^{\prime} &  & \,\lambda_{1\zeta}^{0} &  & \,\lambda_{{21}} &
& \,\lambda_{{3}}^{\prime}\\
\,\lambda_{{2}}^{\prime} &  & \lambda_{2\zeta}^{0} &  & \lambda_{{3}}^{\prime}
&  & \,\lambda_{{22}}%
\end{array}
\right\vert >0\,, \label{vac}%
\end{equation}
where \,$\lambda_{x}^{0}\equiv\mathrm{Min}(0,\lambda_{x})$, \,$\lambda_{{1}%
}^{\prime}=\min(0,\lambda_{{31}}+\lambda_{{41}})$, \,$\lambda_{{2}}^{\prime
}=\min(0,\lambda_{{32}}+\lambda_{{42}})$,\, and \,$\lambda_{{3}}^{\prime}%
=\min(0,\lambda_{{6}}+\lambda_{{7}})$.\, In addition, for the theory to remain
perturbative the magnitudes of the $\lambda$ parameters need to be capped.
Thus, in numerical work we impose \,$|\lambda_{x}|<8\pi$\, for the individual
couplings, which is similar to the condition in the two-Higgs-doublet
case\,\,\cite{Kanemura:1999xf}.

A complementary limitation on $\lambda_{x}$ comes from the demand that the amplitudes for
the scalar-scalar scattering \,$s_1^{}s_2^{}\to s_3^{}s_4^{}$\, at high energies respect
tree-level unitarity.
Also consequential is to ensure that the scalar couplings have values that maintain
the vanishing of the VEVs of the inert doublets.
We elaborate on these extra restrictions in Appendix\,\,\ref{theoretical}.
Numerically, they turn out to be rather mild.

\subsection{Electroweak precision tests\label{ewpd}}

\begin{figure}[t]
\includegraphics[width=0.63\textwidth]{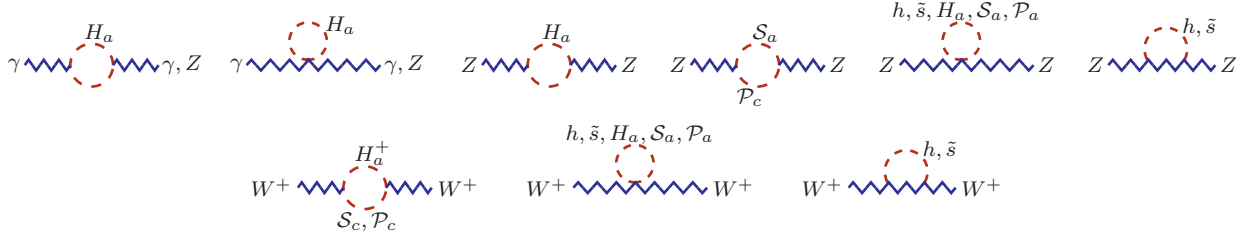} \vspace{-1ex}%
\caption{Feynman diagrams for the contributions of the scalar singlet and
inert scalar doublets to the oblique electroweak parameters $\Delta S$ and
$\Delta T$.}%
\label{diagrams}%
\end{figure}

The nonstandard interactions in Eq.\,(\ref{Lgauge}) and those induced by the
$\phi$-$\varsigma$ mixing bring about changes, $\Delta S$ and $\Delta T$, to
the so-called oblique electroweak parameters $S$ and $T$ which encode the
impact of new physics not coupled directly to SM fermions \cite{Peskin:1991sw}%
. At the one-loop level \cite{Peskin:1991sw,pdg}
\begin{align}
\frac{\alpha\Delta S}{4c_{\mathrm{w}}^{2}s_{\mathrm{w}}^{2}}  &  =
\frac{A_{ZZ}\left(  m_{Z}^{2}\right)  -A_{ZZ}(0)}{m_{Z}^{2}}-A_{\gamma\gamma
}^{\prime}(0) - \frac{c_{\mathrm{w}}^{2}-s_{\mathrm{w}}^{2}}{c_{\mathrm{w}%
}s_{\mathrm{w}}}\,A_{\gamma Z}^{\prime}(0)\,,\nonumber\label{ST}\\
\alpha\Delta T  &  = \frac{A_{WW}(0)}{m_{W}^{2}}-\frac{A_{ZZ}(0)}{m_{Z}^{2}%
}\,,
\end{align}
where $A_{\mathtt{XY}}(q^{2})$ are functions that can be extracted from the
vacuum polarization tensors \,$\Pi_{\mathtt{XY}}^{\mu\nu}(q^{2}%
)=A_{\mathtt{XY}}(q^{2})g^{\mu\nu}+[q^{\mu}q^{\nu}\;$terms]\, of the SM gauge
bosons due to the new scalars' impact at the loop level, and $A_{\mathtt{XY}%
}^{\prime}(0)=[dA_{\mathtt{XY}}(q^{2})/dq^{2}]_{q^{2}=0}$. Here the pertinent
loop diagrams are depicted in Fig.\,\ref{diagrams}.

After evaluating them and subtracting the SM contributions, we arrive at
\begin{align}
6\pi\,\Delta S  &  \,=\, \ln\frac{m_{\mathcal{S}_{1}}m_{\mathcal{S}_{2}}%
}{m_{H_{1}}m_{H_{2}}}-\frac{5}{6}+\mathrm{cos}^{2}(\theta_{S}+\theta
_{P})\,\tilde{F}\big(m_{\mathcal{S}_{1}},m_{\mathcal{P}_{1}},m_{\mathcal{S}%
_{2}},m_{\mathcal{P}_{2}}\big)\nonumber\label{DS}\\
&  ~~~~~+\,\mathrm{sin}^{2}(\theta_{S}+\theta_{P})\,\tilde{F}%
\big(m_{\mathcal{S}_{1}},m_{\mathcal{P}_{2}},m_{\mathcal{S}_{2}}%
,m_{\mathcal{P}_{1}}\big)\nonumber\\
&  ~~~~~+\,s_{\xi}^{2}\Bigg\{\frac{m_{\tilde{s}}^{2}-m_{h}^{2}}{4m_{Z}^{2}%
}+\Bigg[\frac{11m_{Z}^{2}-m_{h}^{2}}{\big(m_{Z}^{2}-m_{h}^{2}%
\big)\raisebox{1pt}{$^2$}}+\frac{1}{m_{Z}^{2}}\Bigg]\frac{\mathcal{F}%
(m_{h},m_{Z})}{2}\nonumber\\
&  \hspace{4em}-\Bigg[\frac{11m_{Z}^{2}-m_{\tilde{s}}^{2}}{\big(m_{Z}%
^{2}-m_{\tilde{s}}^{2}\big)\raisebox{1pt}{$^2$}}+\frac{1}{m_{Z}^{2}%
}\Bigg]\frac{\mathcal{F}(m_{\tilde{s}},m_{Z})}{2}\Bigg\}\,,
\end{align}
\begin{align}
16\pi m_{W}^{2}s_{\mathrm{w}}^{2}\Delta T  &  \;=\;\mathrm{cos}^{2}(\theta
_{H}-\theta_{S})\,\big(\mathcal{F}(m_{H_{1}},m_{\mathcal{S}_{1}}%
)+\mathcal{F}(m_{H_{2}},m_{\mathcal{S}_{2}})\big)\nonumber\\
&  ~~~~~+\;\mathrm{sin}^{2}(\theta_{H}-\theta_{S})\,\big(\mathcal{F}(m_{H_{1}%
},m_{\mathcal{S}_{2}})+\mathcal{F}(m_{H_{2}},m_{\mathcal{S}_{1}}%
)\big)\nonumber\\
&  ~~~~~+\;\mathrm{cos}^{2}(\theta_{H}+\theta_{P})\,\big(\mathcal{F}(m_{H_{1}%
},m_{\mathcal{P}_{1}})+\mathcal{F}(m_{H_{2}},m_{\mathcal{P}_{2}}%
)\big)\nonumber\\
&  ~~~~~+\;\mathrm{sin}^{2}(\theta_{H}+\theta_{P})\,\big(\mathcal{F}(m_{H_{2}%
},m_{\mathcal{P}_{1}})+\mathcal{F}(m_{H_{1}},m_{\mathcal{P}_{2}}%
)\big)\nonumber\\
&  ~~~~~-\;\mathrm{cos}^{2}(\theta_{S}+\theta_{P})\,\big(\mathcal{F}%
(m_{\mathcal{S}_{1}},m_{\mathcal{P}_{1}})+\mathcal{F}(m_{\mathcal{S}_{2}%
},m_{\mathcal{P}_{2}})\big)\nonumber\\
&  ~~~~~-\;\mathrm{sin}^{2}(\theta_{S}+\theta_{P})\,\big(\mathcal{F}%
(m_{\mathcal{S}_{1}},m_{\mathcal{P}_{2}})+\mathcal{F}(m_{\mathcal{P}_{1}%
},m_{\mathcal{S}_{2}})\big)\nonumber\\
&  ~~~~~+\;3s_{\xi}^{2}\big(\mathcal{F}(m_{W},m_{h})-\mathcal{F}%
(m_{W},m_{\tilde{s}})-\mathcal{F}(m_{Z},m_{h})+\mathcal{F}(m_{Z},m_{\tilde{s}%
})\big)\,,
\end{align}
where
\begin{align}
\mathcal{F}(m,n)  &  \;=\;\frac{m^{2}+n^{2}}{2}\,-\,\frac{m^{2}n^{2}}%
{m^{2}-n^{2}}\ln\frac{m^{2}}{n^{2}}\,,\\
& \nonumber\\
\tilde{F}(m_{1},n_{1},m_{2},n_{2})  &  \;=\;\Bigg[\frac{m_{1}^{2}+n_{1}^{2}%
}{\big(m_{1}^{2}-n_{1}^{2}\big)\raisebox{1pt}{$^2$}}-\frac{1}{m_{1}^{2}%
}\Bigg]\frac{\mathcal{F}\big(m_{1},n_{1}\big)}{2}+\Bigg[\frac{m_{2}^{2}%
+n_{2}^{2}}{\big(m_{2}^{2}-n_{2}^{2}\big)\raisebox{1pt}{$^2$}}-\frac{1}%
{m_{2}^{2}}\Bigg]\frac{\mathcal{F}\big(m_{2},n_{2}\big)}{2}\nonumber\\
&  ~~~~+\;\frac{n_{1}^{2}}{4m_{1}^{2}}+\frac{n_{2}^{2}}{4m_{2}^{2}}\,,
\end{align}
and hence \,$\mathcal{F}(m,m)=0$\, and \,$\tilde{F}(m,m,n,n)=5/6$.\, To check
these results, we have also obtained them by employing the formulas provided
in Ref.\thinspace\cite{Grimus:2008nb}. In our numerical analysis, we apply the
$\Delta S$ and $\Delta T$ ranges determined in Ref.\,\cite{pdg}.

\subsection{Collider constraints}

Based on Eq.\,(\ref{Lgauge}), we may infer from the measured widths of the $W
$ and $Z$ bosons and the absence yet of evidence for non-SM particles in their
decay modes that for $a,b=1,2$
\begin{equation}
m_{H_{a}}+m_{\mathcal{S}_{b}}\,>\,m_{W} \,, ~~~~~ m_{H_{a}}+m_{\mathcal{P}%
_{b}}\,>\,m_{W} \,, ~~~~~ 2m_{H_{a}}\,>\,m_{Z}\,, ~~~~~ m_{\mathcal{S}_{a}%
}+m_{\mathcal{P}_{b}} \,>\, m_{Z} \,. \label{WZ}%
\end{equation}
The null results so far of direct searches for new particles at $e^{+}e^{-}$
colliders also translate into lower limits on these masses, especially those
of the charged scalars. In our numerical exploration we will then generally
consider the mass regions \,$m_{H_{a},\mathcal{S}_{a},\mathcal{P}_{a}}>100$\,GeV.\,

Given that the mixing parameter $c_\xi^{}$ defined in Eq.\,(\ref{m2hs}) is the
rescaling factor of the $h$ couplings to ordinary fermions and weak bosons
with respect to their SM counterparts, it needs to satisfy the findings in the
LHC experiments that the $h$ couplings cannot deviate by much more than
{\footnotesize \,$\sim$\,}10\%\, from their SM values\thinspace\thinspace
\cite{atlas+cms}. Moreover, for models with a singlet scalar which mixes with
the noninert scalar doublet, global fits\thinspace\thinspace
\cite{Cheung:2015dta} to the data yield \thinspace$|c_{\xi}%
|\mbox{\footnotesize\,$\gtrsim$\,}0.86$.\, Consequently, we may place the
restraint
\begin{equation}
|s_{\xi}|\,<\,0.3\,.
\end{equation}

Since the decay \,$h\to\gamma\gamma$\, has been measured at the LHC, the data
imply restrictions on the $H_{1,2}$ contributions to $\Gamma(h\rightarrow
\gamma\gamma)$, which we will take into account. On the other hand, although
the invisible decay channel of $h$ is also subject to LHC searches, its limit
will not apply to our case because the inert scalar masses are chosen to
exceed $m_{h}$.

Additional constraints on our scenario come from the fact that searches for
new physics in LHC Run 1 did not produce any clear signals of $\tilde s$ in
its possible decay modes. For the major ones, the data from $pp$ collisions at
\,$\sqrt s=8$\,TeV\, imply the estimated cross-section
limits\,\,\cite{pp2s2gg-1}
\begin{align}
\label{8tevconstraints}\sigma(pp\to\tilde s\to\gamma\gamma)_{8\mathrm{\,TeV}}
\,  &  \lesssim\, 2.3\;\mbox{fb~\cite{Aad:2014ioa,Khachatryan:2015qba}}
\,,\nonumber\\
\sigma(pp\to\tilde s\to\gamma Z)_{8\mathrm{\,TeV}} \,  &  \lesssim\,
4.0\;\mbox{fb~\cite{Aad:2014fha}} \,,\nonumber\\
\sigma(pp\to\tilde s\to WW)_{8\mathrm{\,TeV}} \,  &  \lesssim\,
47\;\mbox{fb~\cite{Aad:2015agg,Khachatryan:2015cwa}} \,,\nonumber\\
\sigma(pp\to\tilde s\to ZZ)_{8\mathrm{\,TeV}} \,  &  \lesssim\,
27\;\mbox{fb~\cite{Aad:2015kna}} \,,\nonumber\\
\sigma(pp\to\tilde s\to hh)_{8\mathrm{\,TeV}} \,  &  \lesssim\,
41\;\mbox{fb~\cite{ATLAS:2014rxa}} \,,\nonumber\\
\sigma(pp\to\tilde s\to t\bar t)_{8\mathrm{\,TeV}} \,  &  \lesssim\,
700\;\mbox{fb~\cite{Chatrchyan:2013lca}} \,.
\end{align}

\section{Constraints on new fermions\label{fdm}}

The interactions of the Dirac singlet fermions $N_k$ with the scalars are
described by Eq.\,(\ref{LN}).
The ${\cal Y}_{1,2}^{}$ terms in ${\cal L}_N$ are responsible for endowing light neutrinos
with masses as well as inducing charged leptons' flavor-violating transitions and anomalous
magnetic moments, all via loop diagrams.
As discussed in Appendix\,\,\ref{Nk}, the $\hat{\texttt Y}_{1,2}$ couplings of $N_k$ in
Eq.\,(\ref{LN}) not only cause their chiral components to mix and transform into Majorana
particles, but also dictate their interactions with $h$ and $\tilde s$.
As this transformation involves mixing matrices with unknown elements and our main purpose
here is to show that the model possesses a viable candidate for DM, in the following for
simplicity we present formulas and results related to $N_k$ where the mixing effects can be
neglected.
Including the latter would only increase the number of free parameters and hence would not
alter our basic conclusions.

\subsection{Radiative neutrino masses and \boldmath$\ell\to\ell'\gamma$ transitions}

The effective Lagrangian for light neutrinos' Majorana masses has the form
\begin{align}
{\cal L}_{m_\nu} \,=\,
-\tfrac{1}{2}\,\overline{\nu_k^{\rm c}}\,({\cal M}_\nu)_{kl\,}^{} P_L^{}\nu_l^{}
\;+\; {\rm H.c.} \,,
\end{align}
where \,$k,l=1,2,3$\, are summed over, \,$P_L=\tfrac{1}{2}(1-\gamma_5^{})$,\,  and
the mass matrix ${\cal M}_\nu$ is related to the neutrino eigenmasses $m_{1,2,3}$ by
the diagonalization formula
\,${\rm diag}\bigl(m_1^{},m_2^{},m_3^{}\bigr)={\cal U}^{\rm T}{\cal M}_\nu\,{\cal U}$\,
involving the Pontecorvo-Maki-Nakagawa-Sakata (PMNS) unitary matrix $\cal U$.
The interactions of the $Z_2$-odd fermions and neutral inert scalars given by Eq.\,(\ref{LN})
provide a mechanism for generating ${\cal M}_\nu$ radiatively via one-loop diagrams
involving $N_j$, ${\cal S}_a$, and ${\cal P}_a$.

Thus, we obtain
\begin{eqnarray} \label{Mnu}
({\cal M}_\nu)_{or}^{} &=&
\frac{({\cal Y}_1^{})_{oj}^{}({\cal Y}_2^{})_{rj}^{} +
({\cal Y}_2^{})_{oj}^{}({\cal Y}_1^{})_{rj}^{}}{16\pi^2}\, M_j^{} \Bigg(
\frac{c_S^{}s_S^{}\,m_{{\cal S}_1}^2}{M_j^2-m_{{\cal S}_1}^2}\ln\frac{m_{{\cal S}_1}^2}{M_j^2}
- \frac{c_S^{}s_S^{}\,m_{{\cal S}_2}^2}{M_j^2-m_{{\cal S}_2}^2}\ln\frac{m_{{\cal S}_2}^2}{M_j^2}
\nonumber \\ && \hspace{32ex} +\;
\frac{c_P^{}s_P^{}\,m_{{\cal P}_1}^2}{M_j^2-m_{{\cal P}_1}^2}\ln\frac{m_{{\cal P}_1}^2}{M_j^2}
- \frac{c_P^{}s_P^{}\,m_{{\cal P}_2}^2}{M_j^2-m_{{\cal P}_2}^2}\ln\frac{m_{{\cal P}_2}^2}{M_j^2}
\Bigg) , ~~~~~
\end{eqnarray}
summation over \,$j=1,2,3$\, being implicit.\footnote{If U(1)$_D$ is unbroken,
this ${\cal M}_\nu$ result becomes that of Ref.\,\cite{Ma:2013yga}, up to an overall minus sign,
with \,$\theta_S=\theta_P=\theta$\, and \,$m_{{\cal S}_a}=m_{{\cal P}_a}=m_{\chi_a}$\,
as defined therein.}
One notices that the ${\cal M}_\nu$ elements are identically zero if one of ${\cal Y}_{1,2}^{}$
is absent or \,$s_S^{}=s_P^{}=0$,\, implying that the presence of both $\eta_{1,2}^{}$ is
necessary for creating the masses of light neutrinos.
However, \,$({\cal M}_\nu)_{kl}^{}=0$\, can still happen if
\,$m_{{\cal S}_1}=m_{{\cal S}_2}$\, and \,$m_{{\cal P}_1}=m_{{\cal P}_2}$\, simultaneously.

The $Z_2$-odd fermions and $H_a^\pm$ together give rise to one-loop diagrams responsible for
\,$\ell\to\ell'\gamma$\, transitions which are subject to stringent experimental constraints.
The diagrams lead us to the branching fraction of the flavor-violating decay
\,$\ell_r\to\ell_o\gamma$\,
\begin{eqnarray} \label{l2l'g}
{\cal B}(\ell_r\to\ell_o\gamma) \,=\,
\frac{3\alpha\,{\cal B}(\ell_r\to\ell_o\nu\bar\nu)\,v^4}{32\pi} \left|
\raisebox{2pt}{\footnotesize$\displaystyle\sum_{j=1,2,3}$} \Bigg[
\frac{{\mathbb C}_{ojrj}^{}}{m_{H_1}^2}\; \mathbb{F}\Bigg(\frac{M_j^2}{m_{H_1}^2}\Bigg) +
\frac{{\mathbb C}_{ojrj}'}{m_{H_2}^2}\; \mathbb{F}\Bigg(\frac{M_j^2}{m_{H_2}^2}\Bigg)
\Bigg] \right|^2
\end{eqnarray}
and a contribution to the anomalous magnetic moment $a_\mu$ of the muon
\begin{eqnarray} \label{g-2}
\delta a_\mu^{} \,=\, \frac{-m_\mu^2}{16\pi^2}\,
\raisebox{2pt}{\footnotesize$\displaystyle\sum_{j=1,2,3}$} \Bigg[
\frac{{\mathbb C}_{2j2j}^{}}{m_{H_1}^2}\; \mathbb{F}\Bigg(\frac{M_j^2}{m_{H_1}^2}\Bigg) +
\frac{{\mathbb C}_{2j2j}'}{m_{H_2}^2}\; \mathbb{F}\Bigg(\frac{M_j^2}{m_{H_2}^2}\Bigg) \Bigg] \,,
\end{eqnarray}
where \,$\mathbb{F}(x)=\big(1-6x+3x^2+2x^3-6x^2\,\ln x\big)/\big[6(1-x)^4\big]$,\,
\begin{align}
{\mathbb C}_{ojrj}^{} &\,=\, c_H^2\big({\cal Y}_1^{}\big)_{oj} \big({\cal Y}_1^*\big)_{rj}
+ s_H^2\big({\cal Y}_2^{}\big)_{oj} \big({\cal Y}_2^*\big)_{rj} \,,
\nonumber \\
{\mathbb C}_{ojrj}' &\,=\, s_H^2\big({\cal Y}_1^{}\big)_{oj}\big({\cal Y}_1^*\big)_{rj}
+ c_H^2\big({\cal Y}_2^{}\big)_{oj} \big({\cal Y}_2^*\big)_{rj} \,.
\end{align}
Since \,$0\le\mathbb{F}(x)\le1/6$\, for \,$x\ge0$,\, it is obvious from the last two equations
that the contribution of the $Z_2$-odd particles in this model to $a_\mu^{}$ is never positive,
\,$\delta a_\mu^{}\le0$.\,

Experiments have indicated that neutrino masses are tiny and that the room for new physics in
\,$\ell\to\ell'\gamma$\, transitions continues to shrink.
One can then see from Eqs.\,\,(\ref{l2l'g}) and (\ref{g-2}) that the elements of the Yukawa
coupling matrices ${\cal Y}_{1,2}^{}$ generally cannot be sizable, unless $M_j$ are very large,
$\theta_{S,P}$ are small, or fine cancellations occur.

\subsection{Fermionic dark matter}

We select the lightest mass eigenstate among the singlet fermions to be lighter than all other
$Z_2$-odd particles, and so it is a candidate for DM.
There are many final states into which it can annihilate, depending on its mass, such as
\,$\ell_o^{-}\ell_r^{+}, \nu_o^{}\nu_r^{}, q\bar{q},W^{+}W^{-}, ZZ, hh,h\tilde{s},
\tilde{s}\tilde{s}$,\, where $q$ is a\,\,quark.
The $\ell_o^{-}\ell_r^{+}$ and $\nu_o^{}\nu_r^{}$ modes, mostly due to $t$- and $u$-channel
diagrams mediated by the inert scalars, are controlled by $({\cal Y}_{1,2}^{})_{j1}^{}$.
Although these couplings are not big, they can bring about consequential contributions
to the DM annihilation rate, as will be addressed later.
Also potentially pertinent are contributions involving the $b\bar b,W^{+}W^{-}, ZZ$, and
$t\bar t$ final-states and arising at tree level from $h$- and $\tilde s$-exchange diagrams in
the $s$ channel, which depend on the other Yukawa couplings, $(\hat{\texttt Y}_{1,2})_{11}^{}$.
The cross sections of some of these processes are relegated to Appendix\,\,\ref{Nk}.

Given that direct searches for DM have led to stringent restrictions on the DM interaction
with the nucleon, we need to take them into account.
In this case, the DM-nucleon scattering proceeds largely from $t$-channel diagrams mediated
at tree level by $h$ and $\tilde s$.
The resulting cross-section is also written down in Appendix\,\,\ref{Nk}.

\section{Numerical results\label{numres}}

Since the $\tilde{s}$ couplings to SM fermions and weak bosons are $s_{\xi}$
times their SM Higgs counterparts, we can estimate the gluon fusion and
vector-boson ($W$ and $Z$) fusion contributions to the cross section
$\sigma(pp\rightarrow\tilde{s})$ from those of a 750\,GeV SM Higgs at
\,$\sqrt{s}=13$\thinspace TeV,\thinspace\ namely\thinspace\thinspace
\cite{cs8tev,cs13tev} \thinspace$\sigma_{g\mathrm{F}}\big(pp\rightarrow
h_{\mathrm{SM}}^{750}\big)
=4.693\times156.8\;\mathrm{fb}\simeq736\;$fb\, and $\sigma_{\mathrm{VBF}%
}\big(pp\rightarrow h_{\mathrm{SM}}^{750}\big)
=2.496\times52.35\;\mathrm{fb}\simeq131\;$fb.\, Thus
\begin{equation}
\sigma_{g\mathrm{F+VBF}}(pp\rightarrow\tilde{s})\,=s_{\xi}^{2}\times
867~\mathrm{fb}\,. \label{cspp2s1}%
\end{equation}
Another contribution to $\sigma(pp\rightarrow\tilde{s})$ comes from photon
fusion\thinspace\thinspace\cite{Jaeckel:2012yz,Csaki:2015vek,
Fichet:2015vvy,Csaki:2016raa}. It has been considered in other studies on this
diphoton excess~\cite{gammaF} and can expectedly yield substantial effects if
$\mathcal{B}(\tilde{s}\rightarrow\gamma\gamma)$ is sizable. At $\,\sqrt
{s}=13\,$TeV,\, the cross section of this production mode is\thinspace
\thinspace\cite{Csaki:2016raa}
\begin{equation}
\sigma_{\gamma\mathrm{F}}(pp\rightarrow\tilde{s})\,=\,1.08\times10^{4}%
\times\frac{\Gamma_{\tilde{s}}}{45~\mathrm{GeV}}\times\mathcal{B}(\tilde
{s}\rightarrow\gamma\gamma)~\mathrm{fb}\,, \label{cspp2s2}%
\end{equation}
owing to the elastic, partially inelastic, and fully inelastic collisions of
the protons, the latter two being
dominant\,\,\cite{Fichet:2015vvy,Csaki:2016raa}. Therefore, the total
production cross-section of $\tilde s$ decaying into the diphoton is
\begin{equation}
\label{cs(pp->s->gg)}\sigma(pp\rightarrow\tilde{s}\rightarrow\gamma\gamma)
\,=\,\left[  \sigma_{g\mathrm{F+VBF}}(pp\rightarrow\tilde{s})+\sigma
_{\gamma\mathrm{F}}(pp\rightarrow\tilde{s})\right]  \,\times\mathcal{B}%
(\tilde{s}\rightarrow\gamma\gamma)\,,
\end{equation}
where ${\mathcal{B}}(\tilde{s}\to\gamma\gamma)$ is given by
Eq.\,(\ref{B(s->gg)}).

Employing Eq.\,(\ref{cs(pp->s->gg)}), we explore the parameter space of the
model in order to attain the cross-section level inferred from the ATLAS and
CMS reports on the 750\,GeV diphoton excess\thinspace\thinspace
\cite{atlas:s2gg,cms:s2gg,Aaboud:2016tru,Khachatryan:2016hje},
namely\thinspace\thinspace\cite{pp2s2gg-1}
\begin{equation}
\sigma(pp\rightarrow\tilde{s}\rightarrow\gamma\gamma)_{\mathrm{LHC}}
\,\sim\,(2-13)~\mathrm{fb}\,, \label{csx}%
\end{equation}
as well as the $\tilde s$ total width \,$\Gamma_{\tilde s}\le50$\,GeV.\,
Simultaneously, we take into account the perturbativity, vacuum stability, and
unitarity conditions, oblique electroweak parameter tests, and restraints from LHC
measurements of \,$\mathcal{B}(h\rightarrow\gamma\gamma)$,\, as discussed in
Sec.\,\ref{constraints}. Furthermore, we consider the charged scalars' mass
regions \,$m_{H_{a}}^{2}>100$\,GeV\, and let $\tilde v$, the VEV of the
singlet scalar, vary between 3 and 10\,\,TeV, for \,$\tilde{v}<\mathcal{O}%
$(1\,TeV)\, would be inadequate for helping enhance the $\tilde{s}\gamma
\gamma$ coupling to the right magnitude.

\begin{figure}[!t]
\includegraphics[height=0.31\textwidth,width=0.47\textwidth]{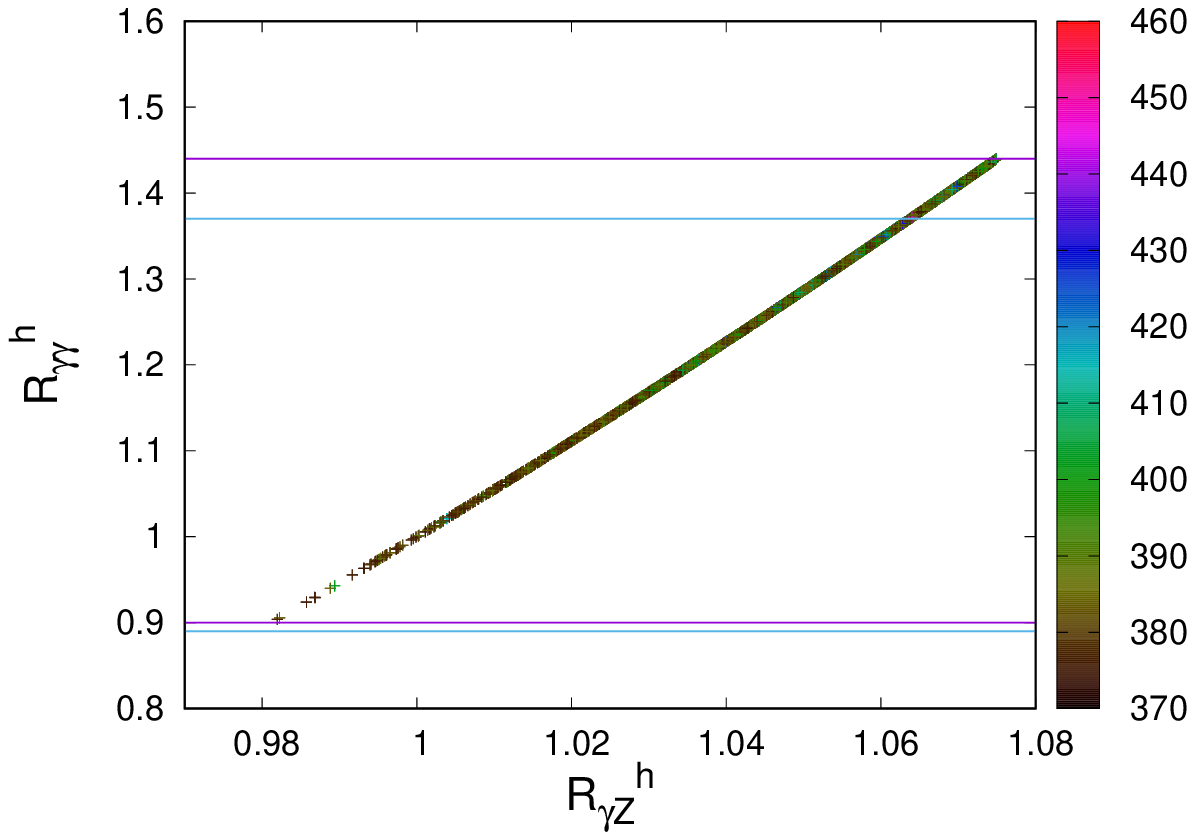}~~%
\includegraphics[height=0.31\textwidth,width=0.47\textwidth]{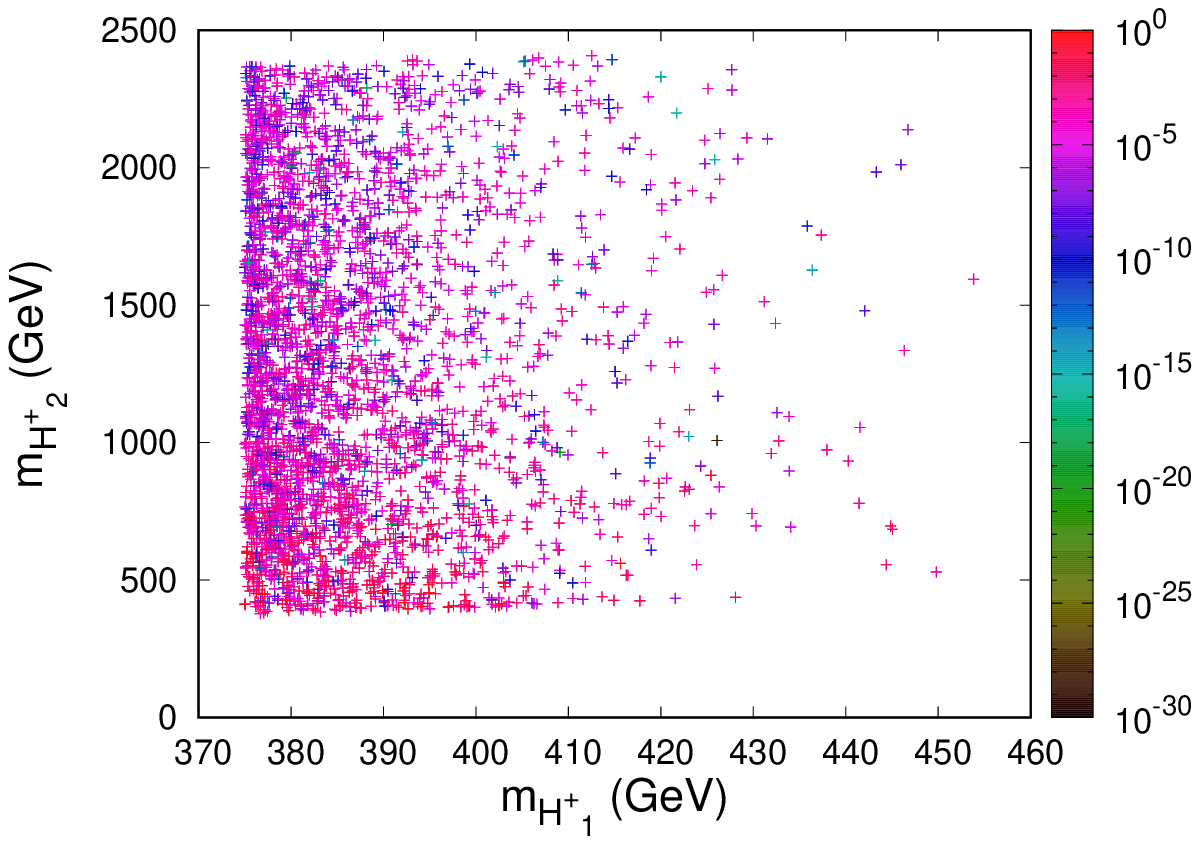}\newline%
\includegraphics[height=0.31\textwidth,width=0.47\textwidth]{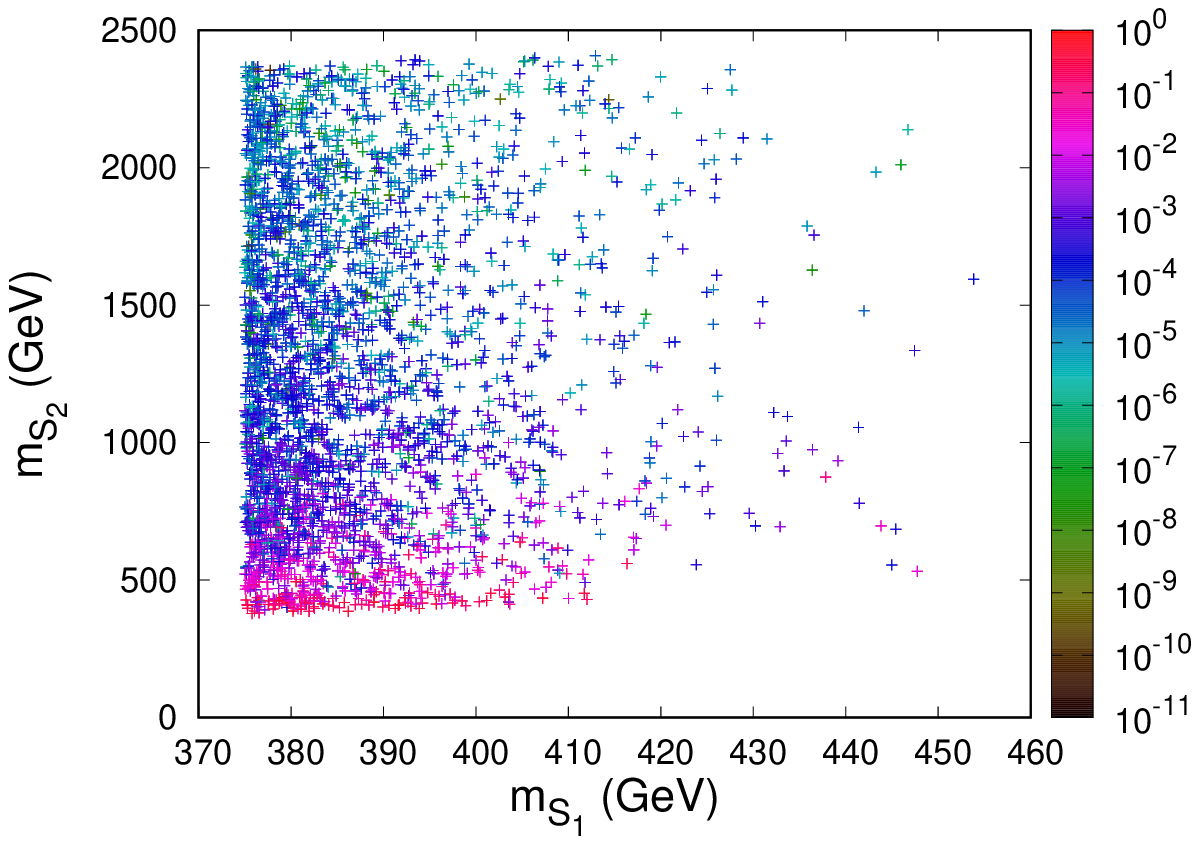}~~%
\includegraphics[height=0.31\textwidth,width=0.47\textwidth]{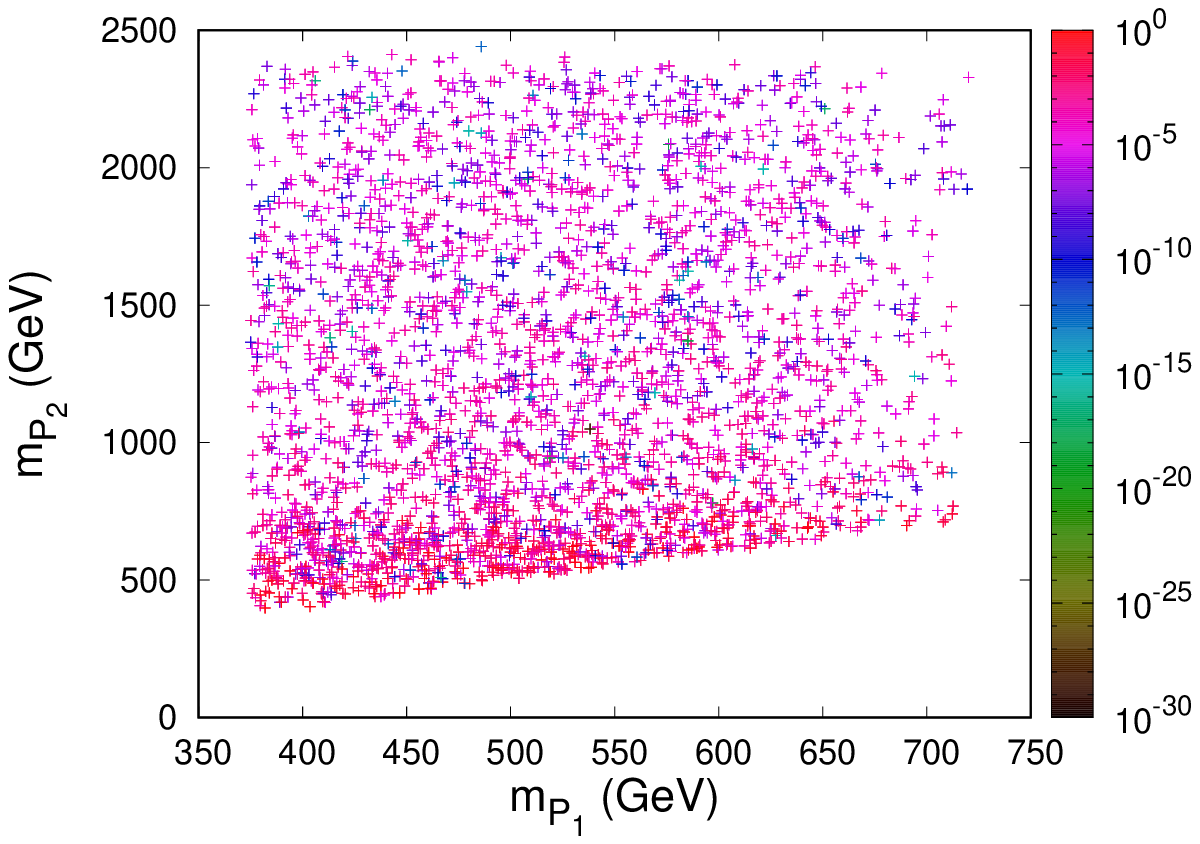}\newline%
\includegraphics[height=0.32\textwidth,width=0.5\textwidth]{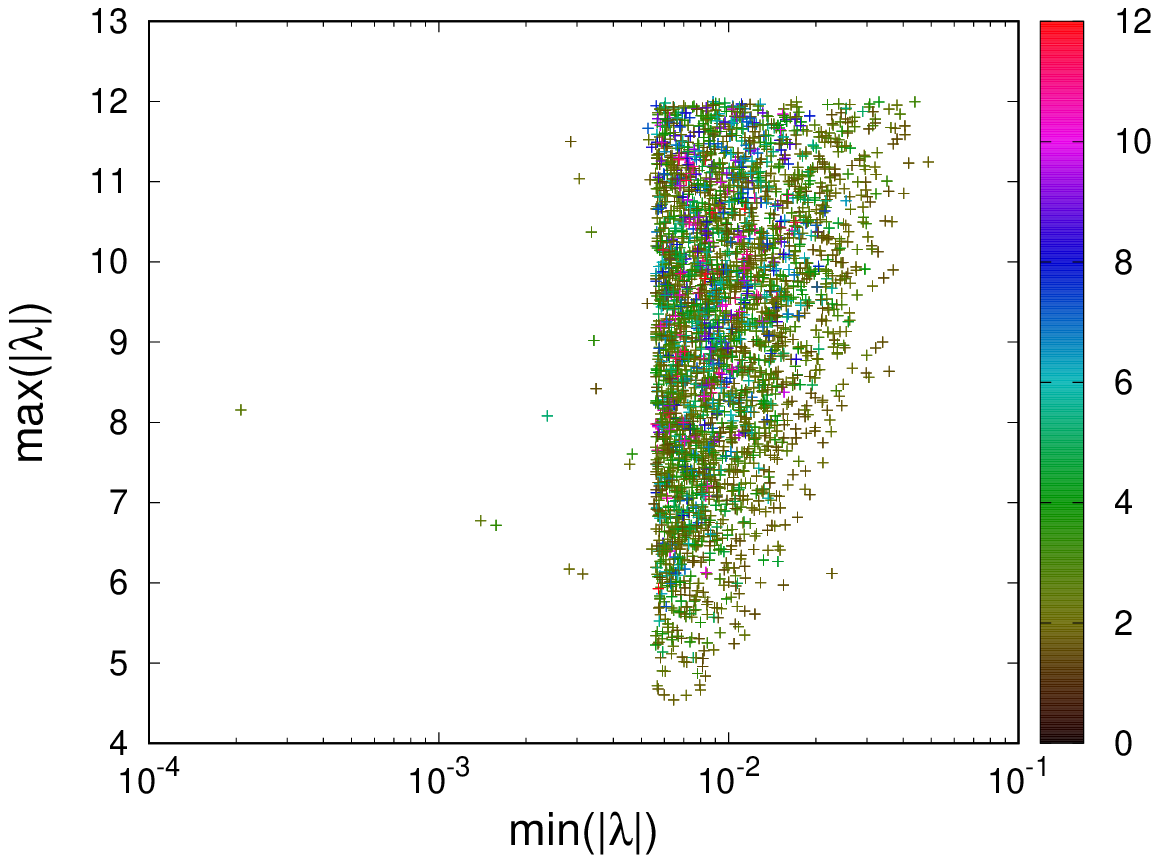}%
\!\!\!\!\includegraphics[height=0.32\textwidth,width=0.5\textwidth]{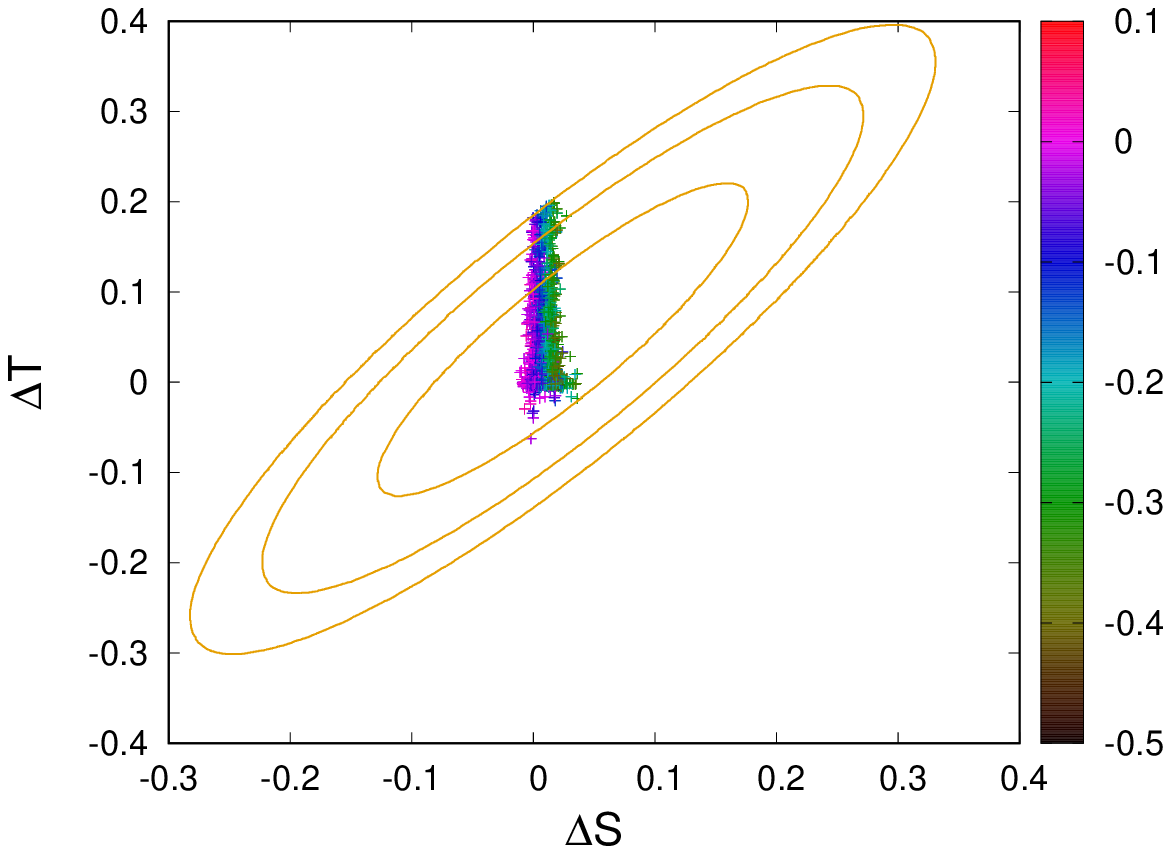}\caption{Top
left panel: the ratio \,$\mathcal{R}_{\gamma\gamma}^{h}=\Gamma(h\rightarrow
\gamma\gamma)/ \Gamma(h\rightarrow\gamma\gamma)_{\textsc{sm}}$\, versus its
$\gamma Z$ counterpart; the horizontal lines mark the 1$\sigma$ ranges of
$\mathcal{R}_{\gamma\gamma}^{h}$ from ATLAS and CMS \cite{atlas+cms}; the
palette reads the value of $m_{H_{1}}$ in GeV. Top right and middle panels:
the masses of the inert scalars; the palettes read their respective mixing
parameters \,$s_{H,S,P}^{2}=\mathrm{sin}^{2}\theta_{H,S,P}$.\, Bottom left
panel: the maximum and minimum magnitudes of the scalars' quartic couplings
$\lambda$s; the palette reads the cross section $\sigma(pp\to\tilde s\to
\gamma\gamma)$ in\,\,fb. Bottom right panel: the oblique electroweak precision
parameters, $\Delta S$ and $\Delta T $; the contours, from smallest to
biggest, represent the empirical 68\%, 95\%, and 99\% confidence level,
respectively; the palette reads the relative mass difference
\,$1-(m_{\mathcal{S}_{1}}+m_{\mathcal{P}_{1}})/(2m_{H_{1}})$.}%
\label{cons}%
\end{figure}

As it turns out, there are viable regions in the model parameter space which
satisfy the different requirements. To illustrate this, we present in
Fig.\thinspace\ref{cons} the distributions of approximately six thousand
randomly-generated benchmark points on the planes of various pairs of
quantities. The top left panel shows \thinspace$\mathcal{R}_{\gamma\gamma}%
^{h}=\Gamma(h\rightarrow\gamma\gamma)/\Gamma_{0}(h\rightarrow\gamma\gamma
)$\thinspace\ versus \thinspace$\mathcal{R}_{\gamma Z}^{h}=\Gamma
(h\rightarrow\gamma Z)/\Gamma_{0}(h\rightarrow\gamma Z)$,\thinspace\ where
$\Gamma_{0}(h\rightarrow\gamma\gamma,\gamma Z)$ stand for the SM rates and are
the same in form as $\Gamma(h\rightarrow\gamma\gamma,\gamma Z)$ in\thinspace
\thinspace Eq.\thinspace(\ref{h2gz}), respectively, but with \thinspace
$c_{\xi}=1$\thinspace\ and \thinspace$\lambda_{hH_{a}H_{a}}=0$.\thinspace
\ Clearly, the model predicts a\thinspace\thinspace positive correlation
between $\mathcal{R}_{\gamma\gamma}^{h}$ and $\mathcal{R}_{\gamma Z}^{h}$,
which will be testable once the empirical information on \thinspace
$h\rightarrow\gamma Z$\thinspace\ has become precise enough. In view of the
purple (blue) horizontal lines marking the 1$\sigma$ range of
$\mathcal{R}_{\gamma\gamma}^{h}$ from ATLAS (CMS) \cite{atlas+cms}, we expect
that many of the predictions which still agree well with the current data will
also be tested by upcoming LHC measurements. In addition, using the color
guide on the vertical palette accompanying the plot, we see that the preferred
values of the mass $m_{H_{1}}$ of the lighter of the charged inert scalars are
not far from \thinspace$m_{\tilde{s}}/2$.\thinspace\ This is not unexpected
because $H_{1}$ with a mass near \thinspace$m_{\tilde{s}}/2$\thinspace\ helps
maximize the \thinspace$\tilde{s}\rightarrow\gamma\gamma$\thinspace\ rate.

The top right and middle panels of Fig.\,\ref{cons} exhibit the distributions
on the $m_{H_{1}}$-$m_{H_{2}}$, $m_{\mathcal{S}_{1}}$-$m_{\mathcal{S}_{2}}$,
and $m_{\mathcal{P}_{1}}$-$m_{\mathcal{P}_{2}}$ planes. Evidently, all the
inert scalars' masses are greater than $m_{\tilde{s}}/2$, but $m_{H_{1}}$, as
already mentioned in the last paragraph, and $m_{\mathcal{S}_{1}}$ do not
reach very far away from\,\,$m_{\tilde{s}}/2$, while $m_{\mathcal{P}_{1}}$ can
go up to 730\,GeV or so. In contrast, the values of $m_{H_{2},\mathcal{S}%
_{2},\mathcal{P}_{2}}$ lie predominantly in the multi-TeV region, but we also
see numerous points corresponding to $m_{H_{2},\mathcal{S}_{2},\mathcal{P}%
_{2}}$ around or below 1\,TeV. For all these masses, the invisible decay
channel of $\tilde s$ into a pair of inert scalars is of course closed. Based
on the accompanying palettes, which provide color guides on the mixing
parameters \,$s_{H,S,P}^{2}=\mathrm{sin}^{2}\theta_{H,S,P}$,\, we deduce that
the mixing in each of the three sectors is very suppressed for the majority of
the benchmark points, with \,$s_{H,S,P}^{2}<10^{-4}$,\, whereas for
\,$m_{H_{2},\mathcal{S}_{2},\mathcal{P}_{2}}%
\,\mbox{\footnotesize$\le$}\,\mathcal{O}$(1\,TeV)\, the mixing can be
significant, with $|s_{H,S,P}|$ as high as $\mathcal{O}$(0.5). Recalling
Eq.\,(\ref{DS}) for $\Delta S$ and $\Delta T$, one realizes that these
different results on the masses and mixing of the inert scalars comply with
the restrictions from electroweak precision data.

The bottom left panel of Fig.\,\ref{cons} depicts the maximum size of
individual quartic scalar couplings versus the minimum size of them, with the
palette reading the cross section $\sigma(pp\to\tilde s\to\gamma\gamma)$
in\,\,fb. It is obvious that one or more of the couplings need to be fairly
large in magnitude, exceeding\,\,6 for most of the benchmarks, which is one of
the conditions for the cross section to rise to the desired level. The
resulting predictions for $\sigma(pp\to\tilde s\to\gamma\gamma)$ appear to lie
primarily within the range of 2-7\,\,fb. We also notice that for a
preponderance of the points the minimum of the quartic couplings is
\mbox{\footnotesize\,$\sim$\,}0.005 or higher, which happens to belong to
$\lambda_{\zeta}$. In these cases, $m_{\tilde s}^{2}$ is chiefly determined by
the \,$m_{\varsigma}^{2}=\lambda_{\zeta}^{}\tilde v^{2}$\, contribution, as
can be concluded from Eq.\,(\ref{m2hs}).

The bottom right panel of Fig.\,\ref{cons} displays the new scalars'
contributions to the oblique electroweak parameters. The plot shows that a
substantial fraction of the benchmarks are within the empirical 1$\sigma$
area. With the palette signifying the amount of relative mass splitting
\,$\hat\delta=1-(m_{\mathcal{S}_{1}}+m_{\mathcal{P}_{1}})/(2m_{H_{1}})$\,
between $H_{1}$ and its lightest neutral counterparts, we also observe that
$\Delta S$ has a dependence on $\hat\delta$, which is similar to the situation
in a newly proposed model involving an inert scalar
doublet~\cite{Ahriche:2016cio}.

Before proceeding to the next figure, we would like to remark that the
aforesaid tendency of the bulk of $m_{H_{2},\mathcal{S}_{2},\mathcal{P}_{2}}$
values to be in the multi-TeV region is attributable to the necessity for one
or more of the scalar quartic couplings and $\tilde{v}$ to be big enough to
boost the \thinspace$\tilde{s}\rightarrow\gamma\gamma$\thinspace\ rate to the
desired amount. On the other hand, $m_{H_{1},\mathcal{S}_{1},\mathcal{P}_{1}}$
have to be fairly close to $m_{\tilde{s}}^{}/2$\thinspace\thinspace and in
numerous cases $m_{H_{2},\mathcal{S}_{2},\mathcal{P}_{2}}$ can also be
sub-TeV, implying that a degree of fine tuning is unavoidable to achieve such
relatively low masses. More precisely, this entails partial cancelation of
order $10^{-4}$ or so mainly between the $\mu_{2a}^{2}$ and $\lambda
_{a\zeta\,}\tilde{v}^{2}$ parts of $m_{c_{a},n_{a}}^{2}$ in $m_{H_{a}%
,\mathcal{S}_{a},\mathcal{P}_{a}}$, as can be inferred from Eqs.\thinspace
\thinspace(\ref{m2H}), (\ref{mSmP}), (\ref{Mc}), and (\ref{m2nz}).

\begin{figure}[b]
\includegraphics[height=0.35\textwidth,width=0.5\textwidth]{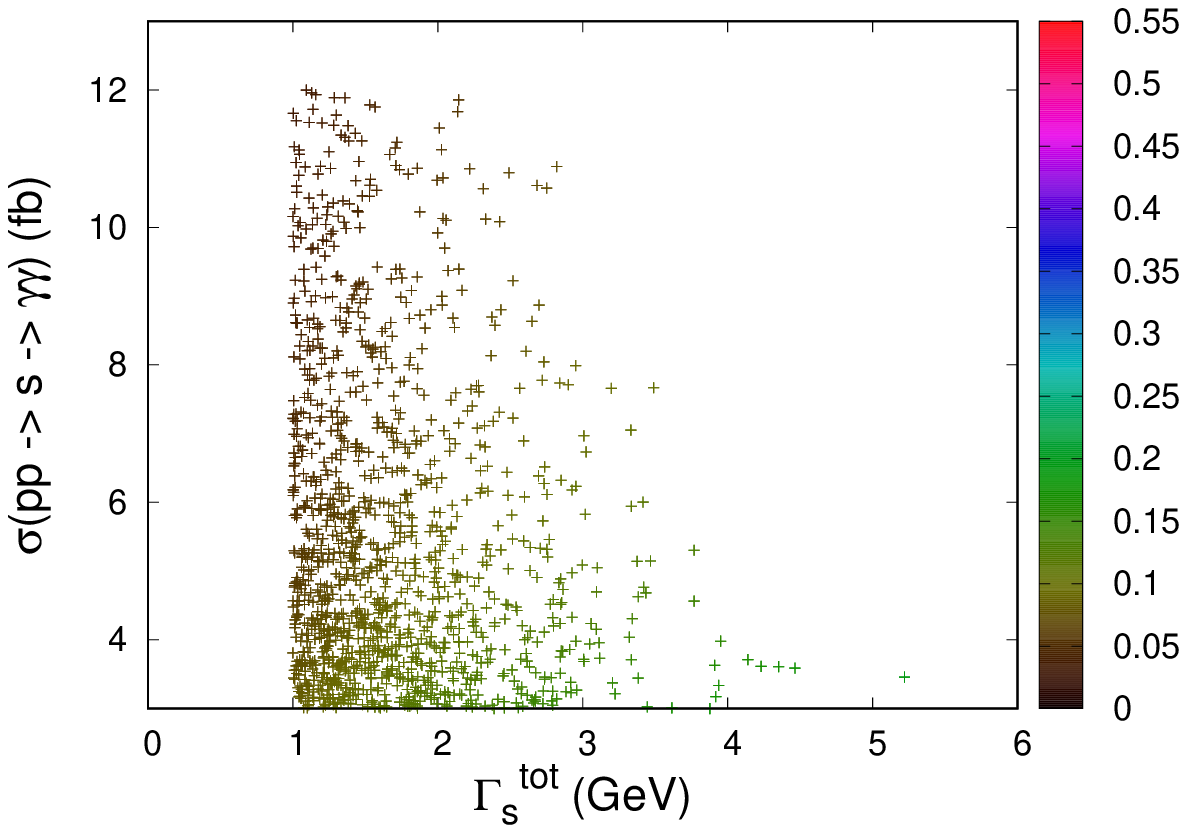}%
\includegraphics[height=0.35\textwidth,width=0.5\textwidth]{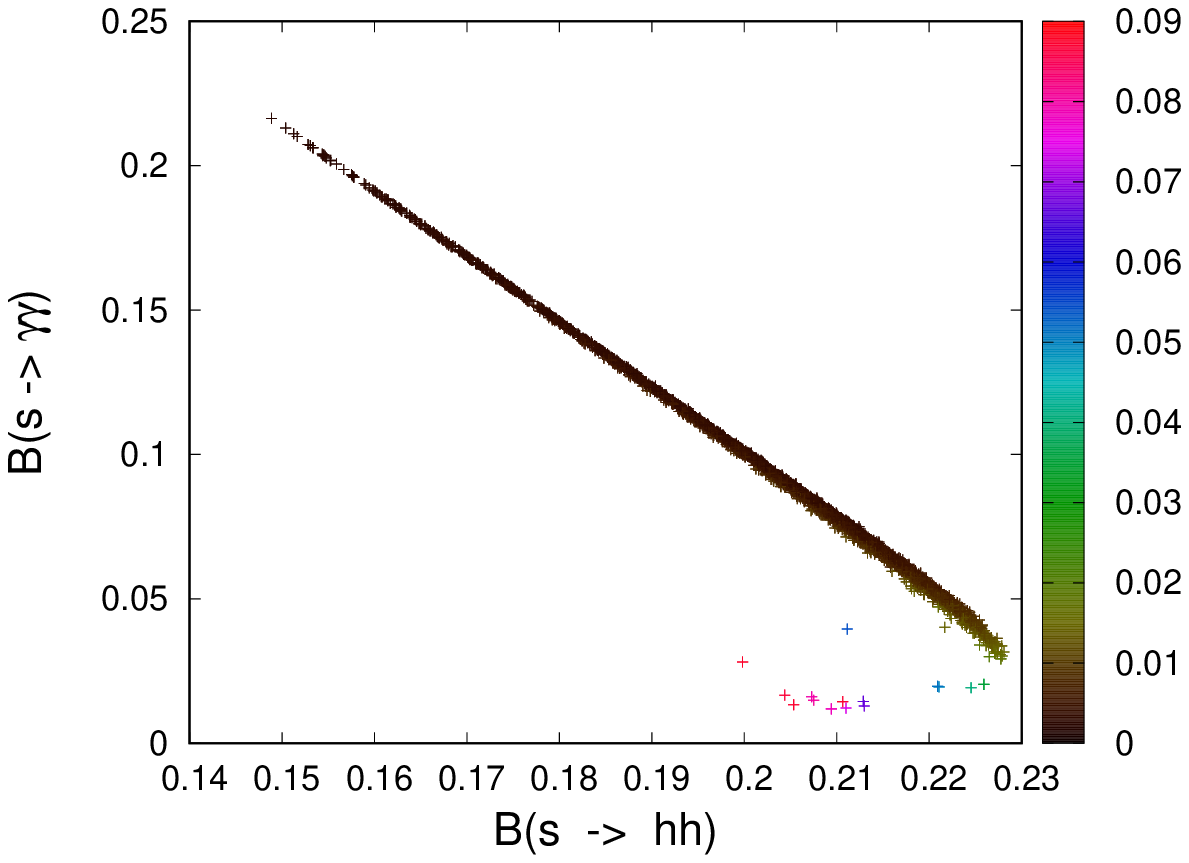}\medskip
\newline%
\includegraphics[height=0.34\textwidth,width=0.53\textwidth]{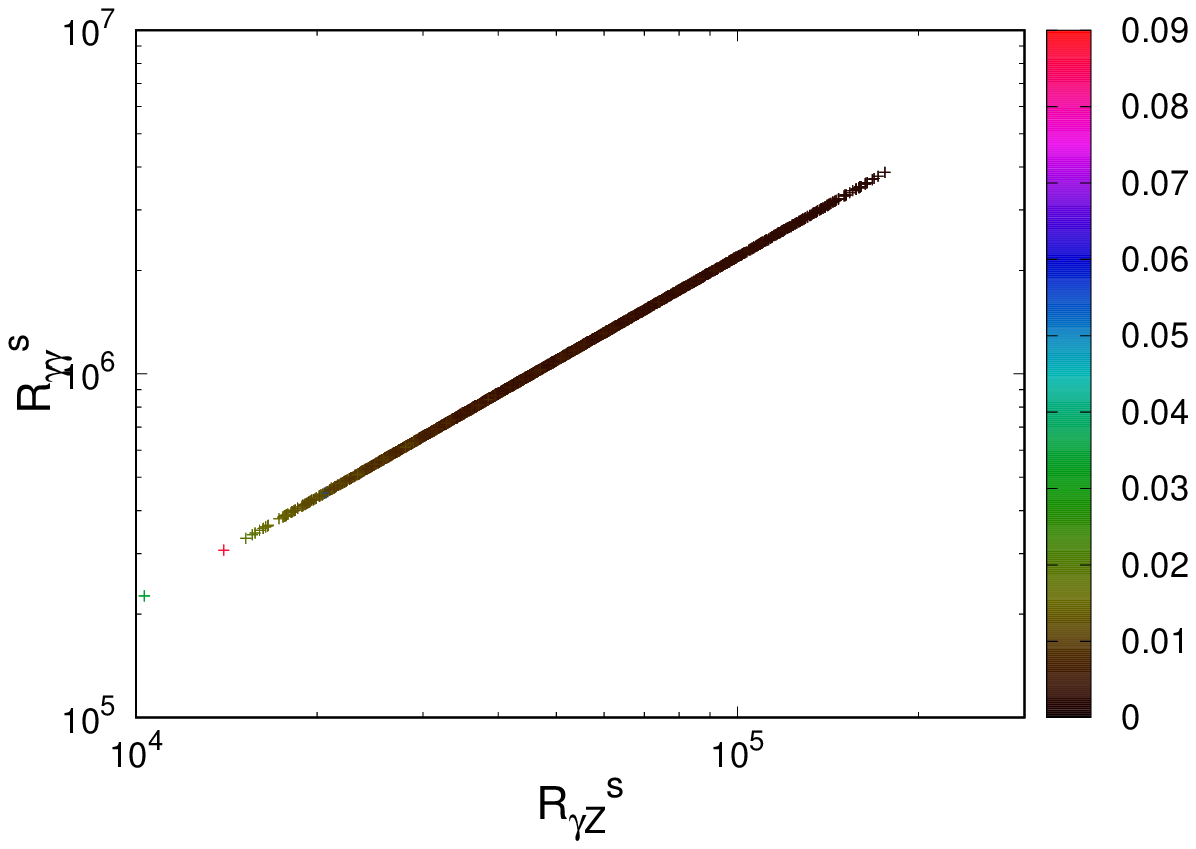}\vspace
{-2ex}\caption{Top left panel: the cross section $\sigma(pp\rightarrow
\tilde{s}\rightarrow\gamma\gamma)$ versus the total $\tilde s $ width
\,$\Gamma_{\tilde{s}}=\Gamma_{\tilde{s}}^{\mathrm{tot}}$;\, the palette reads
the fractional amount of the combined contribution in Eq.\,(\ref{cspp2s1})
from gluon fusion and vector-boson fusion. Top right panel: the branching
fractions $\mathcal{B}(\tilde{s}\rightarrow\gamma\gamma)$ and $\mathcal{B}%
(\tilde{s}\rightarrow hh)$; the palette reads \,$s_{\xi}^{2}=\mathrm{sin}%
^{2}\xi$.\, Bottom panel: the ratios $\mathcal{R}_{\gamma\gamma}^{s}$ and
$\mathcal{R}_{\gamma Z}^{s}$ of $\Gamma(\tilde s\to\gamma\gamma,\gamma Z)$,
respectively, to their counterparts without the $H_{1,2}$ contributions; the
palette reads $s_{\xi}^{2}$. }%
\label{res}%
\end{figure}

For a closer view on $\sigma(pp\to\tilde s\to\gamma\gamma)$, we graph
benchmarks for it versus the $\tilde s$ total width, $\Gamma_{\tilde{s}}$, in
the top left panel of Fig.\,\ref{res}. Obviously, our parameter space of
interest can yield a cross section within the empirical range in
Eq.\,(\ref{csx}) and also $\Gamma_{\tilde{s}}$ between {\footnotesize $\sim
$\,}1 and 6\,\,GeV. With the palette reading the fractional value of the
combined contribution from gluon fusion and vector-boson fusion, it is clear
that in these instances the role of photon fusion is crucial, being
responsible for between {\footnotesize $\sim$\,}80 percent and upper-ninety
percent of $\sigma(pp\rightarrow\tilde{s}\rightarrow\gamma\gamma)$.

Correspondingly, as the top right panel of Fig.\,\ref{res} reveals, the
branching fraction $\mathcal{B}(\tilde{s}\rightarrow\gamma\gamma)$ varies from
about 2 to 22 percent, whereas $\mathcal{B}(\tilde{s}\rightarrow hh)$ is
mostly between 15 and 23 percent. The substantial $\mathcal{B}(\tilde
{s}\rightarrow\gamma\gamma)$ numbers have resulted from the aforementioned big
size of one or more of the quartic couplings, $\tilde v$ being in the 3-10 TeV
range, and the $\tilde s\gamma\gamma$ coupling dominated by the $H_{a}$ loop
contribution with \,$m_{H_{1}}\sim m_{\tilde s}/2$\, for the majority of the benchmarks.

Still with the same panel, from the palette one can see that the favored
mixing between the scalar singlet and noninert doublet is rather small, with
\,$s_{\xi}^{2}\le\mathcal{O}$(0.02),\, which is expected at least on account
of the requirements from electroweak precision data and compatible with
results found in very recent literature\,\,\cite{Falkowski:2015swt}.
Accordingly, one can deduce from Eq.\,(\ref{m2hs}), for \,$m_{\phi}%
^{}<m_{\varsigma}^{}$,\, the approximation \,$m_{\tilde s}^{}\sim\sqrt
{\lambda_{\zeta}}\,\tilde v$,\, and for our choice of \,$\tilde{v}%
=3$-10\,\,TeV\, this causes $\lambda_{\zeta}$ to be quite suppressed, below
$\mathcal{O}$(0.06). These findings fit the comments earlier concerning the
bottom left panel of Fig.\,\ref{cons}.

The bottom panel of Fig.\,\ref{res} depicts some comparison of \,$\tilde
s\to\gamma\gamma$\, and \,$\tilde s\to\gamma Z$,\, particularly \,$\mathcal{R}%
_{\gamma\gamma}^{\tilde s}=\Gamma(\tilde s\to\gamma\gamma)/ \Gamma_{0}(\tilde
s\to\gamma\gamma)$\, versus \,$\mathcal{R}_{\gamma Z}^{\tilde s}=\Gamma(\tilde
s\to\gamma Z)/\Gamma_{0}(\tilde s\to\gamma Z)$,\, with $\Gamma_{0}(\tilde
s\to\gamma\gamma,\gamma Z)$ being the same in form as $\Gamma(\tilde
s\to\gamma\gamma,\gamma Z)$ in Eq.\,(\ref{s2gz}), respectively, but with
\,$\lambda_{hH_{a}H_{a}}=0$.\, The graph reveals that the $H_{1,2}$ loops can
enhance the \,$\tilde s\to\gamma\gamma,\gamma Z$\, rates by several orders of
magnitude relative to the case without $H_{1,2}$ and that there is
a\,\,positive correlation between $\Gamma(\tilde s\to\gamma\gamma,\gamma Z)$,
which can be checked experimentally. In addition, for all the benchmarks our
computation yields \,$\mathcal{R}_{\gamma\gamma}^{\tilde s}\simeq
22\,\mathcal{R}_{\gamma Z}^{\tilde s}$\, as well as \,$\Gamma(\tilde
{s}\rightarrow\gamma\gamma)\simeq1.3\,\Gamma(\tilde{s}\rightarrow\gamma Z)$.\,
The latter translates into \,$\sigma(pp\to\tilde s\to\gamma\gamma
)\simeq1.3\,\sigma(pp\to\tilde s\to\gamma Z)$\, and hence constitutes another
signature of the model which may also be checked soon at the LHC, as the
prediction for $\sigma(pp\to\tilde s\to\gamma Z)$ is roughly an order of
magnitude below the upper limits recently reported by
ATLAS~\cite{Aaboud:2016trl} and CMS~\cite{CMS:2016ypt}.

Other signatures may be accessible by probing \,$pp\to\tilde s\to
hh,W^{+}W^{-},ZZ,t\bar t$,\, although their cross sections depend on $\xi$ and
other parameters. Still, given that \,$\mathcal{B}(\tilde{s}\rightarrow
hh)=\mathcal{O}$(0.2)\, as indicated above, the $hh$ channel is potentially
reachable if the $h$ pair can be observed with good precision. Furthermore,
since \,$\tilde s\to W^{+}W^{-},ZZ,t\bar t$\, have rates adhering to the ratio
in Eq.\,(\ref{rateratio}), the cross sections of \,$pp\to\tilde s\to
W^{+}W^{-},ZZ,t\bar t$\, are predicted to obey the same ratio. Therefore, if
\,$|s_{\xi}|=\mathcal{O}$(0.1),\, they may be sufficiently measurable to allow
us to test these predictions.

\begin{figure}[t]
\medskip\includegraphics[width=0.55\textwidth]{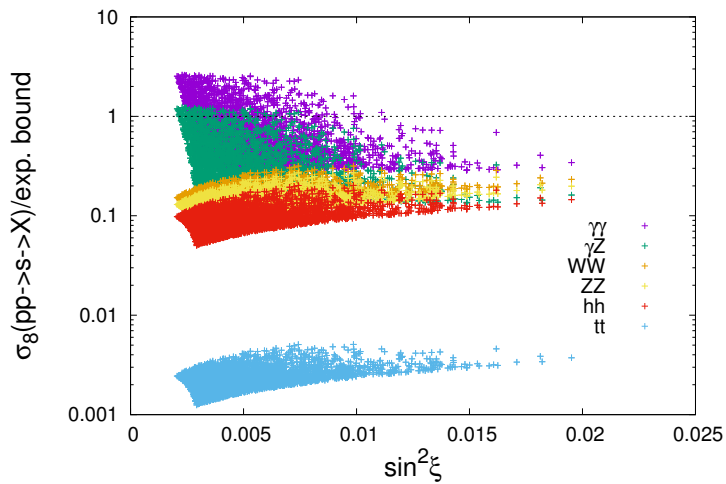} \vspace{-2ex}
\caption{The cross sections of \,$pp\rightarrow\tilde{s}\rightarrow\mathsf{X}
$\, for \,$\mathsf{X}=\gamma\gamma,\gamma Z,W^{+}W^{-},ZZ,hh,t\bar t$\, at the
center-of-mass energy \,$\sqrt s=8$\,TeV,\, each divided by the
corresponding experimental upper-bound from Eq.\,(\ref{8tevconstraints}), versus
the $\phi$-$\varsigma$ mixing parameter $s_{\xi}^{2}$. Points above the
horizontal dotted line are excluded.}%
\label{cs8tev}%
\end{figure}

Now, analogously to Eq.\,(\ref{cs(pp->s->gg)}), at \,$\sqrt s=8$\,TeV\, the
cross section of \,$pp\rightarrow\tilde{s}$\, is
\begin{align}
\sigma(pp\rightarrow\tilde{s})_{8\mathrm{\,TeV}}^{} \,  &  =\, \sigma
_{g\mathrm{F+VBF}}^{}(pp\rightarrow\tilde{s})_{8\mathrm{\,TeV}}^{} \,+\,
\sigma_{\gamma\mathrm{F}}^{}(pp\rightarrow\tilde{s})_{8\mathrm{\,TeV}}^{}%
\end{align}
consisting of the gluon-, vector-boson-, and photon-fusion
contributions\,\,\cite{cs8tev,Csaki:2016raa}
\begin{align}
\sigma_{g\mathrm{F+VBF}}^{}(pp\rightarrow\tilde{s})_{8\mathrm{\,TeV}}^{} \,
&  =\, s_{\xi}^{2}\times(156.8+52.35)\mathrm{fb} \,,\nonumber\\
\sigma_{\gamma\mathrm{F}}^{}(pp\rightarrow\tilde{s})_{8\mathrm{\,TeV}}^{} \,
&  =\, 5.5\times10^{3}\times\frac{\Gamma_{\tilde{s}}}{45~\mathrm{GeV}}%
\times\mathcal{B}(\tilde{s}\rightarrow\gamma\gamma)~\mathrm{fb} \,.
\label{gammaF8tev}%
\end{align}
Using these, we can evaluate \,$\sigma_{8}(pp\rightarrow\tilde{s}%
\rightarrow\mathsf{X})\equiv\sigma(pp\rightarrow\tilde{s})_{8\mathrm{\,TeV\,}%
}^{}\mathcal{B}(\tilde s\rightarrow\mathsf{X})$\, in relation to $s_{\xi}^{2}$
for \,$\mathsf{X}=\gamma\gamma,\gamma Z,W^{+}W^{-},ZZ,hh,t\bar t$\, divided by
the corresponding experimental limits in Eq.\,(\ref{8tevconstraints}). We
display the results in Fig.\,\ref{cs8tev}. It is evident from this plot that
these restraints lead to a\,\,significant decrease in the number of viable
points, as a high percentage of the $\gamma\gamma$ benchmarks (in red) resides
above the horizontal dotted line. However, currently there is considerable
uncertainty in the ratio between the 8 TeV and 13 TeV estimates of the
photon-fusion contributions, which could imply a reduction of the first
numerical factor in $\sigma_{\gamma\mathrm{F}}(pp\rightarrow\tilde
{s})_{8\mathrm{\,TeV}}$ in Eq.\,(\ref{gammaF8tev}) by up to twice or
more\thinspace\thinspace\cite{Fichet:2015vvy,Csaki:2015vek,Csaki:2016raa}. As
a consequence, a substantial portion of the parameter space represented by our
scan points may evade the no-signal constraints from the Run\,1 searches.

Finally, we would like to illustrate how the new fermions $N_j$ via
the calculated quantities in Eqs.\,(\ref{Mnu})-(\ref{g-2}) and Appendix\,\,\ref{Nk} are
subject to the available experimental data on neutrino masses, the \,$\ell\to\ell'\gamma$\,
decays, and the muon anomalous magnetic moment $a_\mu$.
Assuming $N_1$ to be the DM candidate, we take into account as well the constraints from
the observed relic abundance and DM direct searches.
We employ specifically the results of a recent fit to global neutrino data~\cite{nudata},
\,${\cal B}(\mu\to e\gamma)<4.2\times10^{-13}$\, from the MEG experiment~\cite{meg},
\,${\cal B}(\tau\to\mu\gamma)<4.4\times10^{-8}\;$ and the relic density value
\,$\Omega\hat h^2=0.1186\pm0.0020$\, from the Particle Data Group~\cite{pdg}, and the newest
upper-limit on DM-nucleon spin-independent elastic cross-section set by the LUX
Collaboration~\cite{lux}.
Thus, in the same numerical scans as before, we let ${\cal Y}_a^{}$,
$\big(\hat{\texttt Y}_a^{}\big){}_{11}^{}$, and $M_j$ vary within the ranges
\,$-0.5\le\mathcal{Y}_a^{}\le0.5$,\,
\,$\big|\big(\hat{\texttt Y}_1^{}+\hat{\texttt Y}{}_2^*\big){}_{11}^{}\big|\le1$,\,
and \,$M_j=1$-375\,\,GeV,\, with $M_{2,3}$ and also the inert scalar masses being chosen
to exceed $1.1\,M_1$ to avoid coannihilation effects.

\begin{figure}[t]
\includegraphics[height=0.35\textwidth,width=0.49\textwidth]{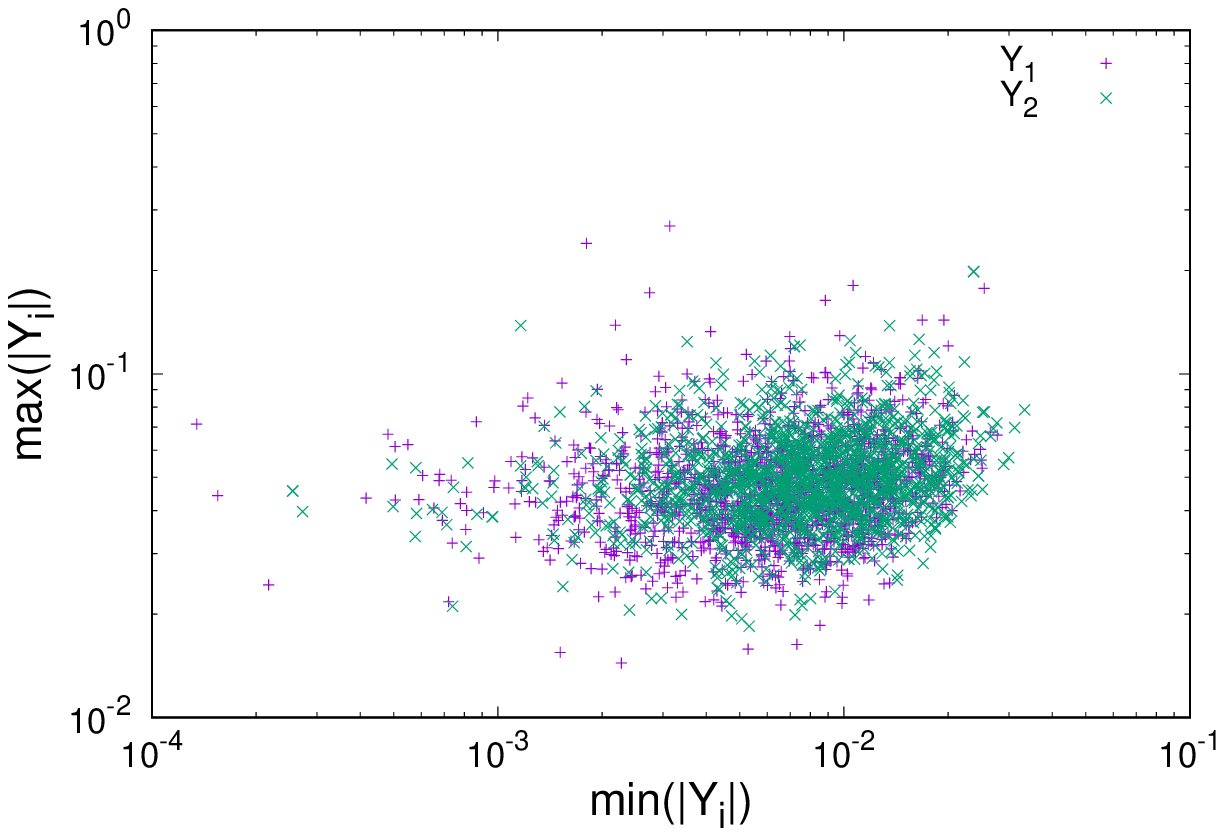}~~~%
\includegraphics[height=0.35\textwidth,width=0.49\textwidth]{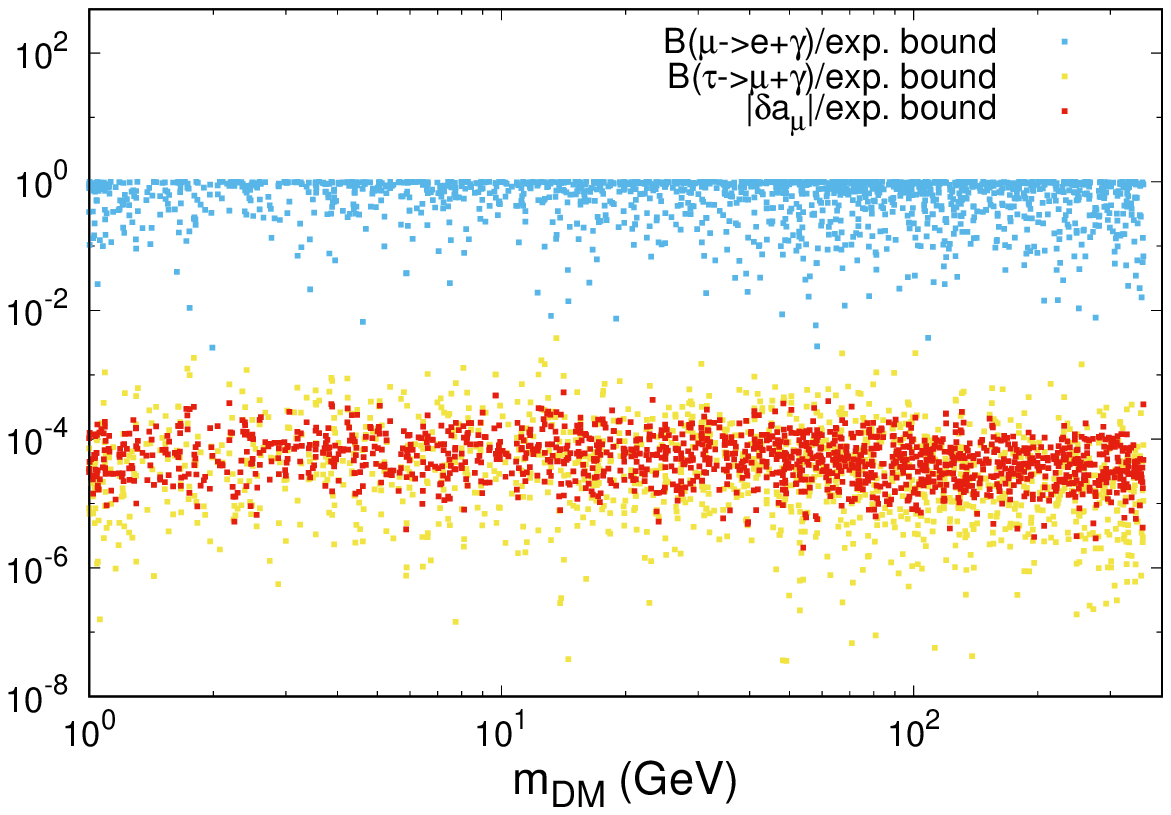}\vspace{-2ex}
\caption{Left: the maximal and minimal values of $|\mathcal{Y}_{1,2}^{}|$
subject to constraints from neutrino oscillation, lepton-flavor violation, and DM data.
Right: the predicted ${\cal B}(\mu\to e\gamma)$, ${\cal B}(\tau\to\mu\gamma)$, and
$|\delta a_\mu|$, normalized by their respective experimental upper-bounds,
versus the DM mass, \,$m_{\rm DM}^{}=M_1^{}$.}
\label{LFV}
\end{figure}

\begin{figure}[t]
\includegraphics[height=0.35\textwidth,width=0.5\textwidth]{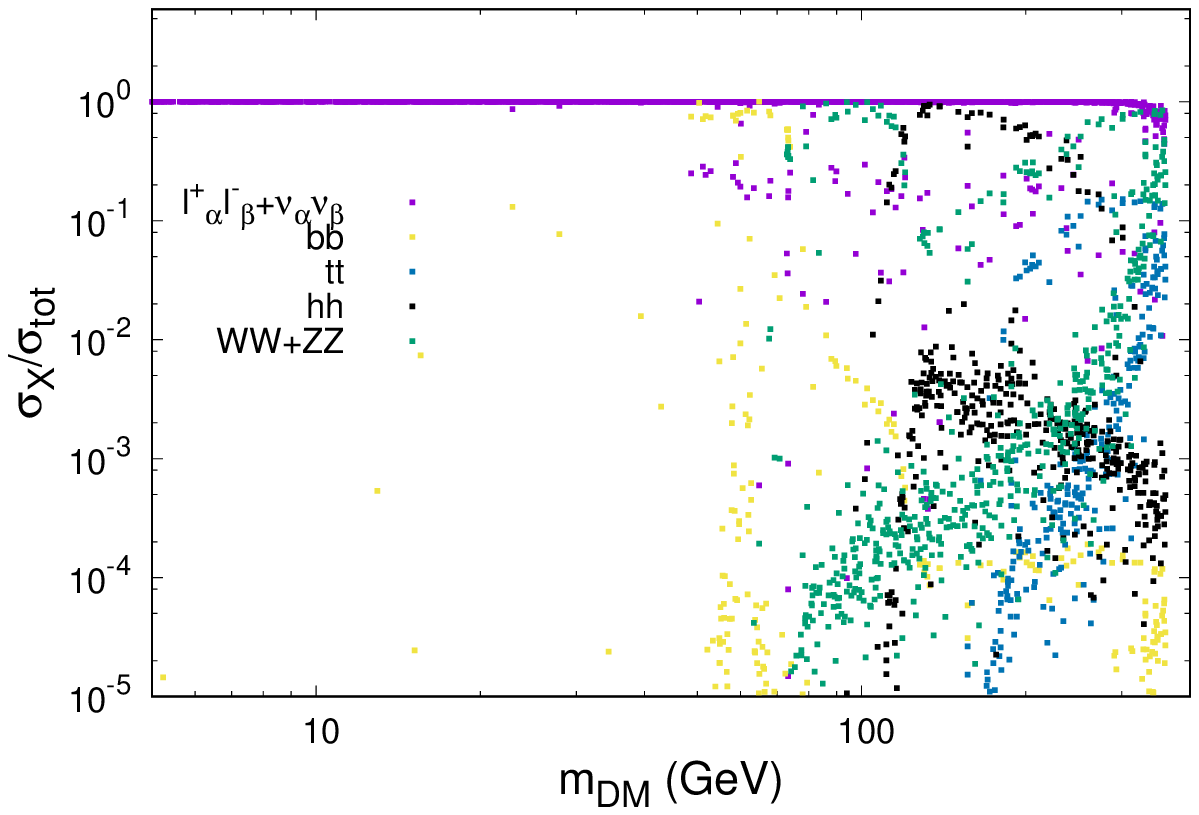}%
\includegraphics[height=0.35\textwidth,width=0.5\textwidth]{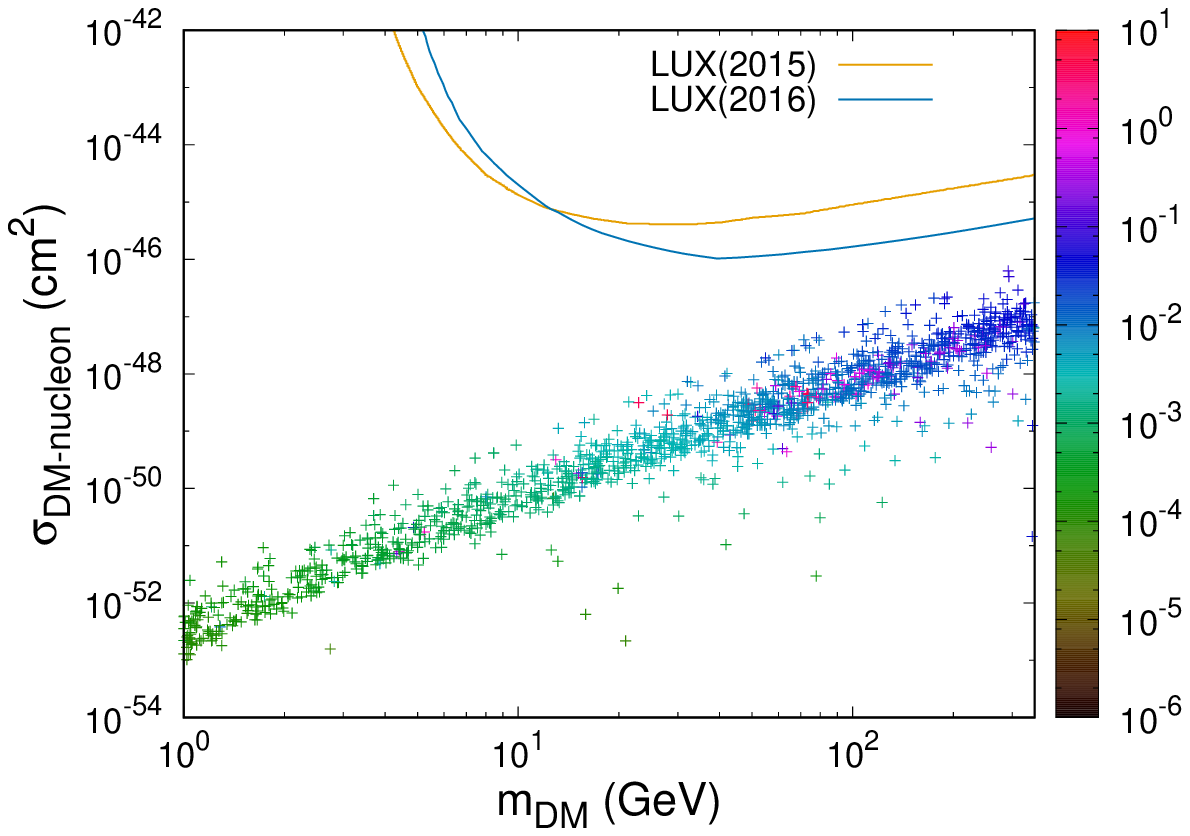}\vspace{-2ex}
\caption{Left: the relative contribution of \,$N_1N_1\to X$\, to the DM-annihilation total
cross-section corresponding to the observed relic density at freeze-out.
Right: the predicted DM-nucleon cross-section versus the DM mass, $M_1$, compared to
the newest LUX upper-bound~\cite{lux}; the palette reads the magnitude of
$\big(\hat{\texttt Y}{}_1^{}+\hat{\texttt Y}{}_2^*\big){}_{11}^{}$.}
\label{DM}
\end{figure}

In Fig.\,\ref{LFV} we display the results for the absolute values of $\mathcal{Y}_a^{}$ and
for ${\cal B}(\mu\to e\gamma)$, ${\cal B}(\tau\to\mu\gamma)$, and $|\delta a_\mu^{}|$
versus the DM mass, $M_1$.
Given that $\delta a_\mu^{}$ in Eq.\,(\ref{g-2}) is not positive and that
the measured and SM values of $a_\mu^{}$ presently differ by
\,$a_\mu^{\rm exp}-a_\mu^{\textsc{sm}}=(288\pm80)\times10^{-11}$~\cite{pdg},
in the scans we have required $|\delta a_\mu|$ to be less than the one-sigma error in
this difference, \,$|\delta a_\mu^{}|<8\times10^{-10}$.\,
In\,\,Fig.\,\ref{DM}, the left panel depicts the relative contributions of the major
DM-annihilation channels \,$N_1N_1\to X$\, to the total annihilation rate that satisfies
the relic density requirement.
Evidently, the $\ell\bar\ell'$ and $\nu\nu'$ contributions, which involve the ${\cal Y}_{1,2}$
elements, are substantial or dominant in the chosen DM mass range, although the other channels,
which are controlled by the $\hat{\texttt Y}_{1,2}$ elements, can also be important in
different mass regions.
The right panel of Fig.\,\ref{DM} shows that the predicted DM-nucleon cross-section is well
below the latest limit from LUX~\cite{lux}, as $(\hat{\texttt Y}{}_{1,2})_{11}$ can be small
enough.
Clearly, there is still ample room in the model parameter space that is compatible with
the existing data.

\section{Conclusions\label{conclusion}}

In this work we have considered the possibility that the observed diphoton
excess at an invariant mass of about 750\thinspace\thinspace GeV recently
reported by the ATLAS and CMS Collaborations is an indication of a new
spinless particle.
To explain it, we propose an extension of the SM with a new sector comprising
two inert scalar doublets, $\eta_{1,2}^{}$, one scalar singlet, $\zeta$, and three Dirac
singlet fermions, $N_{1,2,3}$, all of which transform under a dark Abelian gauge symmetry,
\textrm{U(1)}$_{D}$. We identify $\tilde{s}$, the heavier one
of the mass eigenstates from the mixing of the singlet with the noninert
doublet, as the 750\thinspace GeV resonance. The inert doublets play an
indispensable role because their charged components $H_{1,2}^{\pm}$ can give
rise to the loop-induced $\tilde{s}\gamma\gamma$ coupling of the right
strength, with suitable choices of the model parameters and without the
inclusion of extra colored fermions or bosons.
The presence of both inert doublets is also crucial because their components and the singlet
fermions together are responsible for endowing light neutrinos with radiative mass.
These new scalar doublets and fermions are odd under a $Z_2$ symmetry which naturally
emerges after the spontaneous breaking of \textrm{U(1)}$_{D}$.
We choose the lightest mass eigenstate among the singlet fermions to be the lightest $Z_2$-odd
particle and consequently it can serve as a candidate for DM.

After taking into account the perturbativity condition, the vacuum stability bound, and
the constraints from electroweak precision tests, we show that within the allowed parameter
space the production cross-section $\sigma(pp\rightarrow\tilde{s}\rightarrow
\gamma\gamma)$ can be of order a few fb, mainly due to the sizable
contribution from photon fusion in our scenario, while the total width
$\Gamma_{\tilde{s}}$ lies in the range of 1-6\thinspace\thinspace GeV. The
upcoming data from the LHC with improved precision can be expected to test
this prediction for $\Gamma_{\tilde{s}}$.
In addition, we point out that the model also predicts roughly similar cross-sections of
\thinspace$pp\rightarrow\tilde{s}\rightarrow\gamma\gamma,\gamma Z$\thinspace\
and a specific ratio involving the cross-sections of
\thinspace$pp\rightarrow\tilde{s}\rightarrow W^{+}W^{-},ZZ,t\bar{t}$,\thinspace\ all of
which may be experimentally verified in the near future.
Lastly, we demonstrate that the interactions of the new fermions can be made to fulfill
the restraints from neutrino mass, lepton-flavor violation, muon $g$$-$2, and DM data.

As a final note, after this work was submitted for publication, the ATLAS and CMS Collaborations
reported~\cite{lhc0} that their 2016 data with four times larger statistics than those analyzed
in their earlier reports~\cite{Aaboud:2016tru,Khachatryan:2016hje}
revealed no significant diphoton excess above the SM backgrounds at around 750 GeV.
Although this does not necessarily rule out the existence of a heavy diphoton resonance,
such a particle if existent would have a relatively smaller production cross-section and hence
probably require much more statistics to discover.
On the other hand, theoretically this implies that it would likely be easier for our model of
interest to accommodate the particle, as the scalar couplings and singlet VEV would not need
to have the big values seen in our scans.
Moreover, as the model parameter space is still considerable, if there is another tentative hint
of a heavy diphoton resonance in the future, significantly improved empirical constraints on
the various observables discussed above would be needed to probe the model extensively.

\acknowledgments

A.A. is supported by the Algerian Ministry of Higher Education and Scientific
Research under the CNEPRU Project No. D01720130042. He would like to thank
Xiao-Gang He and J.T. for the warm hospitality at NTU-CTS, where this work was
initiated. The work of G.F. was supported in part by the research grant
NTU-ERP-102R7701. The work of J.T. was supported in part by the MOE Academic
Excellence Program (Grant No. 102R891505). We gratefully acknowledge partial
support from the National Center for Theoretical Sciences of Taiwan.

\appendix

\section{Scalar masses and couplings\label{app}}

In Eq.\thinspace(\ref{L2}), the squared-mass matrices $M_{\phi\varsigma
,\mathcal{C},0}^{2}$ and column matrix $\eta_{0}$ are given by
\begin{equation}
M_{\phi\varsigma}^{2} \,= \left(\begin{array}[c]{ccc}%
m_{\phi}^{2} && \displaystyle\frac{m_{\phi\varsigma}^{2}}{2} \vspace{5pt} \\
\displaystyle\frac{m_{\phi\varsigma}^{2}}{2} && m_{\varsigma}^{2}
\end{array}\right)  ,~~~~~
m_{\phi}^{2}\,=\,\lambda_{1}v^{2}\,,~~~~m_{\varsigma}%
^{2}\,=\,\lambda_{\zeta}\tilde{v}^{2}\,,~~~~m_{\phi\varsigma}^{2}%
\,=\,2\lambda_{3\zeta}^{~~\;}v\tilde{v}\,,\label{M0}%
\end{equation}%
\begin{equation}
M_{\mathcal{C}}^{2} \,= \left(\begin{array}[c]{ccc}
m_{c_{1}}^{2} && \displaystyle\frac{m_{c\zeta}^{2}}{2} \vspace{5pt} \\
\displaystyle\frac{m_{c\zeta}^{2\ast}}{2} && m_{c_{2}}^{2}
\end{array}\right)  ,~~~~~
m_{c_{a}}^{2}\,=\,\mu_{2a}^{2}+\frac{\lambda_{3a}v^{2}%
+\lambda_{a\zeta}\tilde{v}^{2}}{2}\,,~~~~m_{c\zeta}^{2}\,=\,\sqrt{2}%
\,\mu_{\eta\zeta\,}\tilde{v}\,,\label{Mc}
\end{equation}
\begin{equation}
M_{0}^{2} \,= \left(\begin{array}[c]{ccccccc}%
m_{n_{1}}^{2} && \displaystyle\frac{m_{n\zeta}^{2}}{2} && 0 &&
\displaystyle\frac{-\mathrm{Im}\,m_{c\zeta}^{2}}{2} \vspace{5pt} \\
\displaystyle\frac{m_{n\zeta}^{2}}{2} && m_{n_{2}}^{2} &&
\displaystyle\frac{\mathrm{Im}\,m_{c\zeta}^{2}}{2} && 0 \vspace{5pt}\\
0 && \displaystyle\frac{\mathrm{Im}\,m_{c\zeta}^{2}}{2} && m_{n_{1}}^{2} &&
\displaystyle -\frac{\tilde{m}_{n\zeta}^{2}}{2} \vspace{5pt} \\
\displaystyle \frac{-\mathrm{Im}\,m_{c\zeta}^{2}}{2} && 0 &&
\displaystyle -\frac{\tilde{m}_{n\zeta}^{2}}{2} && m_{n_{2}}^{2}
\end{array}\right)  ,~~~~~~~
\eta_{0} \,= \left(\begin{array}[c]{c}
\mathrm{Re}\,\eta_{1}^{0} \vspace{5pt} \\ \mathrm{Re}\,\eta_{2}^{0} \vspace{5pt} \\
\mathrm{Im}\,\eta_{1}^{0} \vspace{5pt} \\ \mathrm{Im}\,\eta_{2}^{0} \end{array}\right)  ,
\end{equation}
\begin{align}
m_{n_{a}}^{2}  & \,=\, \mu_{2a}^{2}+\frac{\big(\lambda_{3a}+\lambda_{4a}%
\big)v^{2}+\lambda_{a\zeta}\tilde{v}^{2}}{2}\,=\,m_{c_{a}}^{2}+\frac{\lambda_{4a}v^{2}}{2} \,,
\vphantom{|_{\int_\int^|}^{}} \nonumber \\
m_{n\zeta}^{2} & \,=\, \frac{\lambda_{5}}{2}\,v^{2}+\mathrm{Re}\,m_{c\zeta}^{2} \,, ~~~~~~~~~~
\tilde{m}_{n\zeta}^{2} \,=\,
\frac{\lambda_{5}}{2}\,v^{2} - \mathrm{Re}\,m_{c\zeta}^{2}\,. \label{m2nz}
\end{align}
The constants $\mu_{2a}^{2}$ enter only these mass formulas of the inert
scalars and can be positive or negative if nonzero. To arrive at
$M_{\phi\varsigma}^{2}$ in Eq.\thinspace(\ref{M0}), we have used the
relations
\begin{align} \label{mu1mu2}
\mu_{1}^{2}\,=\,\frac{-\lambda_{1\,}v^{2}-\lambda_{3\zeta\,}\tilde{v}^{2}}{2} \,, ~~~~~~~
\mu_{\zeta}^{2}\,=\,\frac{-\lambda_{\zeta\,}\tilde{v}^{2}-\lambda_{3\zeta\,}v^{2}}{2}\,,
\end{align}
corresponding to the vanishing of the first derivatives of the potential
$\mathcal{V}$ with respect to $\phi$ and $\varsigma$.

If the parameter $\mu_{\eta\zeta} $ in the potential is complex, so is
$M_{\mathcal{C}}^{2}$, which can then be diagonalized with the unitary matrix
$\mathcal{U}_{\mathcal{C}} $ according to
\begin{align}
\label{c11}\mathcal{U}_{\mathcal{C}}^{}  & \,= \left( \begin{array}[c]{cc}
\mathcal{C}_{11}^{} & \mathcal{C}_{12}^{} \vspace{5pt} \\
\mathcal{C}_{21}^{} & \mathcal{C}_{22}^{}
\end{array}\!\right)_{\vphantom{|_{\int}^{\int}}}  , ~~~~~~~
\mathcal{U}_{\mathcal{C}\,}^{\dagger}M_{\mathcal{C}}^{2}\,
\mathcal{U}_{\mathcal{C}}^{} \,=\, \mathrm{diag}\bigl(m_{H_{1}}^{2},_{\,\!}m_{H_{2}}^{2}\bigr) \,,
\nonumber \\
\mathcal{C}_{11}^{} & \,=\, \frac{+\mu_{\eta\zeta} }{\sqrt2\;\big|\mu_{\eta\zeta}
\big|} \sqrt{1+\frac{m_{c_{2}}^{2}-m_{c_{1}}^{2}}{m_{H_{2}}^{2}-m_{H_{1}}^{2}%
}} \,=\, \frac{\mu_{\eta\zeta} }{\big|\mu_{\eta\zeta} \big|\vphantom{|_{\int_|^|}^{}}}
\;\mathcal{C}_{22}^{} \,,
\nonumber\\
\mathcal{C}_{12}^{} & \,=\, \frac{\mu_{\eta\zeta} }{\sqrt2\;\big|\mu_{\eta\zeta}
\big|} \sqrt{1+\frac{m_{c_{1}}^{2}-m_{c_{2}}^{2}}{m_{H_{2}}^{2}-m_{H_{1}}^{2}%
}} \,=\, \frac{-\mu_{\eta\zeta} }{\big|\mu_{\eta\zeta} \big|}\; \mathcal{C}_{21}^{} \,.
\end{align}
As noted earlier, without loss of generality, we can select $\mu_{\eta\zeta}$
to be real, rendering $m_{c\zeta}^{2}$ real as well, in which case
$\mathcal{U}_{\mathcal{C}}$ has the orthogonal form in Eq.\,(\ref{m2H}).

If $\mu_{\eta\zeta} $ is complex, the matrix $\mathcal{O}_{0} $ that
diagonalizes $M_{0}^{2}$ has the form
\begin{align}
\mathcal{O}_{0} \,= \left(\begin{array}[c]{cccc}
\mathcal{N}_{11} & \mathcal{N}_{12} & \mathcal{N}_{13} & \mathcal{N}_{14}
\vspace{5pt}\\
\mathcal{N}_{21} & \mathcal{N}_{22} & \mathcal{N}_{23} & \mathcal{N}_{24}
\vspace{5pt}\\
\mathcal{N}_{31} & \mathcal{N}_{32} & \mathcal{N}_{33} & \mathcal{N}_{34}
\vspace{5pt}\\
\mathcal{N}_{41} & \mathcal{N}_{42} & \mathcal{N}_{43} & \mathcal{N}_{44}%
\end{array}
\right)  ,
\end{align}
where $\mathcal{N}_{rs}$ are mostly complicated. With $\mu_{\eta\zeta} $ being
real instead, these elements are much simpler
\begin{align}
\mathcal{N}_{11}  & \,=\, \frac{\sigma_{11} \;\mathrm{sgn}\big(m_{n\zeta}%
^{2}\big)} {\sqrt2}\sqrt{1+\frac{m_{n_{2}}^{2}-m_{n_{1}}^{2}}{m_{\mathcal{S}%
_{2}}^{2}-m_{\mathcal{S}_{1}}^{2}}} \,, ~~~~ ~~~
\mathcal{N}_{12} \,=\, \frac{\sigma_{12} }{\sqrt2}
\sqrt{1+\frac{m_{n_{1}}^{2}-m_{n_{2}}^{2}}{m_{\mathcal{S}_{2}}^{2}-m_{\mathcal{S}_{1}}^{2}}} \,,
\nonumber\\
\mathcal{N}_{33}  & \,=\, \frac{\sigma_{33} \;\mathrm{sgn}\big(\tilde m_{n\zeta
}^{2}\big)} {\sqrt2}\sqrt{1+\frac{m_{n_{2}}^{2}-m_{n_{1}}^{2}}{m_{\mathcal{P}%
_{2}}^{2}-m_{\mathcal{P}_{1}}^{2}}} _{\vphantom{|}}^{\vphantom{|}} \,, ~~~~~~~
\mathcal{N}_{34} \,=\, \frac{\sigma_{34} }{\sqrt2} \sqrt{1+\frac
{m_{n_{1}}^{2}-m_{n_{2}}^{2}}{m_{\mathcal{P}_{2}}^{2}-m_{\mathcal{P}_{1}}^{2}%
}} \,, \nonumber \\ \vphantom{|_{\int}^{\int}}
\mathcal{N}_{22}  & \,=\, \sigma_{11} \sigma_{12} \,\mathcal{N}_{11} \,, ~~~~~
\mathcal{N}_{21} \,=\, -\sigma_{11} \sigma_{12} \,\mathcal{N}_{12} \,, ~~~~~
\mathcal{N}_{44} \,=\, -\sigma_{33} \sigma_{34} \,\mathcal{N}_{33} \,, ~~~~~
\mathcal{N}_{43} \,=\, \sigma_{33} \sigma_{34} \,\mathcal{N}_{34} \,,
\nonumber\\ \vphantom{|_{\int}^{\int}}
\mathcal{N}_{13}  & \,=\, \mathcal{N}_{14} \,=\, \mathcal{N}_{23} \,=\,
\mathcal{N}_{24} \,=\, \mathcal{N}_{31} \,=\, \mathcal{N}_{32} \,=\,
\mathcal{N}_{41} \,=\, \mathcal{N}_{42} \,=\, 0 \,, \nonumber \\ \vphantom{|^{\int}}
\sigma_{11}^{2}  & \,=\, \sigma_{12}^{2} \,=\, 1 \,, ~~~~~~~ \sigma_{33}^{2}
\,=\, \sigma_{34}^{2} \,=\, 1 \,,
\end{align}
where $\sigma_{11,12,33,34} $ are independent of each other and can each be
either +1 or $-1$, implying that we can choose \,$\sigma_{11} \sigma_{12}
=\sigma_{33} \sigma_{34} =+1$\, to get the form of $\mathcal{O}_{0} $ in
Eq.\,(\ref{mSmP}).

The $\lambda$'s in Eq.\,(\ref{Llambdas}) are
\begin{align} \label{lhhh}
\lambda_{hhh} \,=\, 3c_\xi^{} \left( c_\xi^2 \lambda_1^{}+s_\xi^2\lambda_{3\zeta}^{} \right)
- 3s_\xi^{} \left( c_\xi^2\lambda_{3\zeta}^{} + s_\xi^2\lambda_\zeta^{} \right)
\frac{\tilde{v}}{v} \,, \hspace{7em}
\end{align}
\begin{align}
\label{hHH}\lambda_{hH_{1}H_{1}}  &  = c_{\xi}\left(  c_{H}^{2}\lambda_{31} +
s_{H}^{2}\lambda_{32} \right)  - s_{\xi}\left[  \big(c_{H}^{2}\lambda_{1\zeta}
+ s_{H}^{2}\lambda_{2\zeta} \big) \frac{\tilde{v}}{v}-\sqrt2\,c_{H} s_{H}
\,\frac{\mu_{\eta\zeta} }{v} \right]  ,\nonumber\\
\lambda_{hH_{2}H_{2}}  &  = c_{\xi}\left(  c_{H}^{2}\lambda_{32} + s_{H}%
^{2}\lambda_{31} \right)  - s_{\xi}\left[  \big(c_{H}^{2}\lambda_{2\zeta} +
s_{H}^{2}\lambda_{1\zeta} \big) \frac{\tilde{v}}{v}+\sqrt2\,c_{H} s_{H}
\,\frac{\mu_{\eta\zeta} }{v} \right]  ,
\end{align}
\begin{align}
\label{shh}\lambda_{\tilde{s}hh}  &  = c_{\xi}\left[  \left(  1 - 3s_{\xi}%
^{2}\right)  \lambda_{3\zeta} + 3s_{\xi}^{2} \lambda_{\zeta}\right]  + s_{\xi
}\left[  \left(  1 - 3c_{\xi}^{2}\right)  \lambda_{3\zeta} + 3 c_{\xi}^{2}
\lambda_{1} \right]  \frac{v}{\tilde{v}} \,,
\end{align}
\begin{align}
\label{sHH}\lambda_{\tilde{s}H_{1}H_{1}}  &  = c_{\xi}\left(  c_{H}^{2}%
\lambda_{1\zeta} + s_{H}^{2}\lambda_{2\zeta} - \sqrt2\,c_{H} s_{H} \,
\frac{\mu_{\eta\zeta} }{\tilde{v}} \right)  + s_{\xi}\big(c_{H}^{2}%
\lambda_{31} + s_{H}^{2}\lambda_{32} \big) \frac{v}{\tilde{v}} \,,\nonumber\\
\lambda_{\tilde{s}H_{2}H_{2}}  &  = c_{\xi}\left(  c_{H}^{2}\lambda_{2\zeta} +
s_{H}^{2}\lambda_{1\zeta} + \sqrt2\,c_{H} s_{H} \, \frac{\mu_{\eta\zeta}
}{\tilde{v}} \right)  + s_{\xi}\big(c_{H}^{2}\lambda_{32} + s_{H}^{2}%
\lambda_{31} \big) \frac{v}{\tilde{v}} \,,
\end{align}
\begin{align}
\label{sSS}\lambda_{\tilde{s}{\mathcal{S}}_{1}{\mathcal{S}}_{1}}  &  = c_{\xi
}\left(  c_{S}^{2} \lambda_{1\zeta} + s_{S}^{2} \lambda_{2\zeta} - \sqrt2\,
c_{S} s_{S} \, \frac{\mu_{\eta\zeta} }{\tilde{v}} \right) \nonumber\\
&  +\; s_{\xi}\left[  c_{S}^{2} \left(  \lambda_{31}+\lambda_{41}\right)  +
s_{S}^{2} \left(  \lambda_{32}+\lambda_{42}\right)  - c_{S} s_{S} \,
\lambda_{5} \right]  \frac{v}{\tilde{v}} \,,
\vphantom{|_{\int_|^|} }\nonumber\\
\lambda_{\tilde{s}{\mathcal{S}}_{2}{\mathcal{S}}_{2}}  &  = c_{\xi}\left(
c_{S}^{2} \lambda_{2\zeta} + s_{S}^{2} \lambda_{1\zeta} + \sqrt2\, c_{S} s_{S}
\, \frac{\mu_{\eta\zeta} }{\tilde{v}} \right) \nonumber\\
&  +\; s_{\xi}\left[  c_{S}^{2} \left(  \lambda_{32}+\lambda_{42}\right)  +
s_{S}^{2} \left(  \lambda_{31}+\lambda_{41}\right)  + c_{S} s_{S} \,
\lambda_{5} \right]  \frac{v}{\tilde{v}} \,,
\vphantom{|_{\int_|^|} }\nonumber\\
\lambda_{\tilde{s}{\mathcal{S}}_{1}{\mathcal{S}}_{2}}  &  = c_{\xi}\left[
c_{S} s_{S} \left(  \lambda_{1\zeta} - \lambda_{2\zeta} \right)  +
\big(c_{S}^{2}-s_{S}^{2}\big) \frac{\mu_{\eta\zeta} }{\sqrt2\,\tilde{v}}
\right] \nonumber\\
&  +\; s_{\xi}\left[  c_{S} s_{S} \left(  \lambda_{31}+\lambda_{41}%
-\lambda_{32}-\lambda_{42}\right)  + \tfrac{1}{2}\big(c_{S}^{2}-s_{S}%
^{2}\big) \lambda_{5} \right]  \frac{v}{\tilde{v}} \,,
\end{align}
\begin{align}
\label{sPP}\lambda_{\tilde{s}{\mathcal{P}}_{1}{\mathcal{P}}_{1}}  &  = c_{\xi
}\left(  c_{P}^{2} \lambda_{1\zeta} + s_{P}^{2} \lambda_{2\zeta} + \sqrt2\,
c_{P} s_{P} \, \frac{\mu_{\eta\zeta} }{\tilde{v}} \right) \nonumber\\
&  +\; s_{\xi}\left[  c_{P}^{2} \left(  \lambda_{31}+\lambda_{41}\right)  +
s_{P}^{2} \left(  \lambda_{32}+\lambda_{42}\right)  - c_{P} s_{P} \,
\lambda_{5} \right]  \frac{v}{\tilde{v}} \,,
\vphantom{|_{\int_|^|} }\nonumber\\
\lambda_{\tilde{s}{\mathcal{P}}_{2}{\mathcal{P}}_{2}}  &  = c_{\xi}\left(
c_{P}^{2} \lambda_{2\zeta} + s_{P}^{2} \lambda_{1\zeta} - \sqrt2\, c_{P} s_{P}
\, \frac{\mu_{\eta\zeta} }{\tilde{v}} \right) \nonumber\\
&  +\; s_{\xi}\left[  c_{P}^{2} \left(  \lambda_{32}+\lambda_{42}\right)  +
s_{P}^{2} \left(  \lambda_{31}+\lambda_{41}\right)  + c_{P} s_{P} \,
\lambda_{5} \right]  \frac{v}{\tilde{v}} \,,
\vphantom{|_{\int_|^|} }\nonumber\\
\lambda_{\tilde{s}{\mathcal{P}}_{1}{\mathcal{P}}_{2}}  &  = c_{\xi}\left[
c_{P} s_{P} \left(  \lambda_{1\zeta} - \lambda_{2\zeta} \right)  -
\big(c_{P}^{2}-s_{P}^{2}\big) \frac{\mu_{\eta\zeta} }{\sqrt2\,\tilde{v}}
\right] \nonumber\\
&  +\; s_{\xi}\left[  c_{P} s_{P} \left(  \lambda_{31}+\lambda_{41}%
-\lambda_{32}-\lambda_{42}\right)  + \tfrac{1}{2}\big(c_{P}^{2}-s_{P}%
^{2}\big) \lambda_{5} \right]  \frac{v}{\tilde{v}} \,.
\end{align}
We then find
\begin{align}
\lambda_{hH_{1}H_{1}} + \lambda_{hH_{2}H_{2}}  &  = c_{\xi}\left(
\lambda_{31} + \lambda_{32} \right)  - s_{\xi}\left(  \lambda_{1\zeta} +
\lambda_{2\zeta} \right)  \frac{\tilde{v}}{v} \,,\nonumber\\
\lambda_{\tilde{s}H_{1}H_{1}} + \lambda_{\tilde{s}H_{2}H_{2}}  &  = c_{\xi
}\left(  \lambda_{1\zeta} + \lambda_{2\zeta} \right)  + s_{\xi}\left(
\lambda_{31} + \lambda_{32} \right)  \frac{v}{\tilde{v}} \,.
\end{align}

\section{Conditions for tree-level unitarity and global minimum of potential\label{theoretical}}

One of the consequential restrictions on the parameters in the scalar potential $\cal V$
is that the amplitudes for scalar-scalar scattering \,$s_1^{}s_2^{}\to s_3^{}s_4^{}$\, at high
energies do not violate unitarity.
Analogously to the situation in two-Higgs-doublet models~\cite{Kanemura:1993hm}, for the scalar
pair \,$s_m^{}s_n^{}$\, we can work with the nonphysical components of the scalar doublets and
singlet,
\begin{eqnarray}
\Phi \,= \left(\!\begin{array}{c} \varphi^+ \\ \varphi^0 \end{array}\! \right) , ~~~~
\Phi^\dagger \,=\, \big( \varphi^- ~~~ \varphi^{0*} \big) \,, ~~~~~
\eta_a^{} \,= \left(\!\begin{array}{c} \eta_a^+ \vspace{1pt} \\ \eta_a^0 \end{array}\!\right) ,
~~~~ \eta_a^\dagger \,=\, \big( \eta_a^- ~~~ \eta_a^{0*} \big) \,, ~~~~~
\zeta \,, ~~~ \zeta^* \,.
\end{eqnarray}
Accordingly, we can select the uncoupled sets of orthonormal pairs
\begin{eqnarray} &
\Big\{ \frac{1}{\sqrt2}\,\zeta \zeta \Big\} \,, ~~~ \big\{ \zeta \eta_1^0 \big\} \,, ~~~
\big\{ \zeta \eta_2^{0*} \big\} \,, ~~~ \Big\{ \frac{1}{\sqrt2}\,\eta_1^0 \eta_1^0 \Big\} \,, ~~~
\big\{ \zeta \varphi^0 \big\} \,, ~~~ \big\{ \eta_1^+ \eta_2^-, \eta_1^0 \eta_2^{0*} \big\} \,, ~~~
\Big\{ \frac{1}{\sqrt2}\,\eta_2^{0*} \eta_2^{0*} \Big\} \,,
& \nonumber \\ &
\big\{ \varphi^+ \eta_2^-, \varphi^0 \eta_2^{0*}, \varphi^- \eta_1^+, \varphi^{0*} \eta_1^0
\big\} \,, ~~~
\big\{ \zeta \zeta^*, \varphi^+ \varphi^-, \varphi^0 \varphi^{0*}, \eta_1^+ \eta_1^-,
\eta_1^0 \eta_1^{0*}, \eta_2^+ \eta_2^-, \eta_2^0 \eta_2^{0*} \big\} \,,
& \nonumber \\ &
\big\{ \varphi^+ \eta_1^0, \varphi^0 \eta_1^+ \big\} \,, ~~~~~
\big\{ \varphi^+ \varphi^0, \eta_1^+ \eta_2^0, \eta_2^+ \eta_1^0 \big\} \,, ~~~~~
\big\{ \varphi^+ \eta_2^0, \varphi^0 \eta_2^+ \big\} &
\end{eqnarray}
to construct the matrix containing the tree-level amplitudes for
\,$s_1^{}s_2^{}\to s_3^{}s_4^{}$,\, which at high energies are dominated by the contributions
of four-particle contact diagrams.
We can express the distinct eigenvalues of this matrix as
\begin{eqnarray} &
\lambda_\zeta^{} \,, ~~~ \lambda_{1\zeta}^{} \,, ~~~ \lambda_{2\zeta}^{} \,, ~~~
\lambda_{3\zeta}^{} \,, ~~~ \lambda_{21}^{} \,, ~~~ \lambda_{22}^{} \,, ~~~
\lambda_{31}^{}\pm\lambda_{41}^{} \,, ~~~ \lambda_{32}^{}\pm\lambda_{42}^{} \,, ~~~
\lambda_6^{} \,, ~~~ \lambda_6^{}+2\lambda_7^{} \,, ~~~ \lambda_6^{}-\lambda_7^{} \,,
& \nonumber \\ &
{\cal E}_\pm^{} \,=\, \tfrac{1}{2}(\lambda_{31}+\lambda_{32}+2\lambda_{41}+2\lambda_{42}) \pm
\tfrac{1}{2} \sqrt{(\lambda_{31}-\lambda_{32}+2\lambda_{41}-2\lambda_{42})\raisebox{0.2pt}{$^2$}
+ 9\lambda_5^2} \,,
& \nonumber \\ &
E_\pm^{} \,=\, \tfrac{1}{2}(\lambda_{31}+\lambda_{32})\pm \tfrac{1}{2}
\sqrt{(\lambda_{31}-\lambda_{32})\raisebox{0.2pt}{$^2$}+\lambda_5^2} \,,
& \nonumber \\ &
{\mathbb E}_\pm^{} \,=\, \tfrac{1}{2}(\lambda_1+\lambda_6+\lambda_7)\pm \tfrac{1}{2}
\sqrt{(\lambda_1-\lambda_6-\lambda_7)\raisebox{0.2pt}{$^2$}
+ 2\lambda_5^2} \,, &
\end{eqnarray}
the solutions ${\cal E}_{1,2,3}$ of the cubic polynomial equation
\begin{eqnarray}
0 &=& {\cal E}^3 - \big( \lambda_1^{} + \lambda_{21}^{} + \lambda_{22}^{} \big){\cal E}^2 +
\big[ \lambda_1^{} \big( \lambda_{21}^{} + \lambda_{22}^{} \big) +
\lambda_{21}\lambda_{22} - \lambda_{41}^2 - \lambda_{42}^2 - \lambda_7^2 \big] {\cal E}
\nonumber \\ && +\;
\lambda_1^{} \big( \lambda_7^2 - \lambda_{21}^{}\lambda_{22}^{} \big)
+ \lambda_{21}^{}\lambda_{42}^2+\lambda_{22}^{}\lambda_{41}^2
- 2\lambda_{41}^{}\lambda_{42}^{}\lambda_7^{} \,,
\end{eqnarray}
and the solutions $E_{1,2,3,4}$ of the quartic polynomial equation
\begin{eqnarray}
0 &=& E^4
- \big( 3\lambda_1^{} + 3\lambda_{21}^{} + 3\lambda_{22}^{} + 2\lambda_\zeta^{} \big) E^3
\nonumber \\ && +\;
\big[ \big(9\lambda_1^{}+6\lambda_\zeta^{}\big) \big( \lambda_{21}^{}+\lambda_{22}^{} \big) +
9\lambda_{21}^{}\lambda_{22}^{} - (2\lambda_{31}+\lambda_{41})^2 - (2\lambda_{32}+\lambda_{42})^2
- (2\lambda_6+\lambda_7)^2
\nonumber \\ && ~~~~ +\,
6 \lambda_1^{}\lambda_\zeta^{}
- 2\big(\lambda_{3\zeta}^2+\lambda_{1\zeta}^2+\lambda_{2\zeta}^2\big) \big] E^2
\nonumber \\ && +\;
\big\{ (3\lambda_1+2\lambda_\zeta)\big[(2\lambda_6+\lambda_7)^2-9\lambda_{21}^{}\lambda_{22}^{}\big]
+ \big(6\lambda_{3\zeta}^2-18\lambda_1^{}\lambda_\zeta^{}\big)(\lambda_{21}+\lambda_{22})
+ 6(\lambda_1+\lambda_{21})\lambda_{2\zeta}^2
\nonumber \\ && ~~~~ +\,
6(\lambda_1+\lambda_{22})\lambda_{1\zeta}^2
- \big[ 2(2\lambda_{31}+\lambda_{41})(2\lambda_{32}+\lambda_{42})+4\lambda_{1\zeta}\lambda_{2\zeta}
\big] (2\lambda_6+\lambda_7)
\nonumber \\ && ~~~~ -\,
4 \big[(2\lambda_{31}+\lambda_{41})\lambda_{1\zeta}+(2\lambda_{32}+\lambda_{42})\lambda_{2\zeta}
\big] \lambda_{3\zeta}
+ (3\lambda_{21}+2\lambda_\zeta)(2\lambda_{32}+\lambda_{42})^2
\nonumber \\ && ~~~~ +\,
(3\lambda_{22}+2\lambda_\zeta)(2\lambda_{31}+\lambda_{41})^2 \big\} E
\nonumber \\ && +\;
18 \big[ \lambda_1^{}\big(\lambda_{21}\lambda_{2\zeta}^2-\lambda_{22}^{}\lambda_{1\zeta}^2
+ 3\lambda_{21}^{}\lambda_{22}^{}\lambda_\zeta^{}\big)
- \lambda_{21}^{}\lambda_{22}^{}\lambda_{3\zeta}^2 \big]
+ 2 (2\lambda_6+\lambda_7)^2 \big(\lambda_{3\zeta}^2-3\lambda_1^{}\lambda_\zeta^{}\big)
\nonumber \\ && +\;
12 \big[ \lambda_{21}(2\lambda_{32}+\lambda_{42})\lambda_{2\zeta} +
      \lambda_{22}(2\lambda_{31}+\lambda_{41})\lambda_{1\zeta} \big] \lambda_{3\zeta}
- 4(2\lambda_{31}+\lambda_{41})(2\lambda_{32}+\lambda_{42})\lambda_{1\zeta} \lambda_{2\zeta}
\nonumber \\ && +\;
2 (2\lambda_{31}+\lambda_{41})^2\big(\lambda_{2\zeta}^2-3\lambda_{22}^{}\lambda_\zeta^{}\big)
+ 2 (2\lambda_{32}+\lambda_{42})^2\big(\lambda_{1\zeta}^2-3\lambda_{21}^{}\lambda_\zeta^{}\big)
\nonumber \\ && +\;
4 \big\{ (2\lambda_{31}+\lambda_{41})(2\lambda_{32}+\lambda_{42})\lambda_\zeta
- \big[(2\lambda_{31}+\lambda_{41})\lambda_{2\zeta}+(2\lambda_{32}+\lambda_{42})\lambda_{1\zeta}
\big] \lambda_{3\zeta} \big\} (2\lambda_6+\lambda_7)
\nonumber \\ && +\;
12 \lambda_1 \lambda_{1\zeta} \lambda_{2\zeta} (2\lambda_6+\lambda_7) \,.
\end{eqnarray}
The requirement of unitarity dictates that each of these eigenvalues not exceed $8\pi$
in magnitude.

We now discuss how we ensure that the potential minimum with the VEVs of the inert doublets
being zero is a global minimum.
As usual, we get the possible minima of $\cal V$ from the solutions to
\begin{eqnarray}
\bigg(\frac{\partial {\mathcal{V}}}{\partial b}\bigg)_{b\,
=\langle b\rangle} =\, 0 \,, ~~~~~
\bigg(\frac{\partial{\mathcal{V}}}{\partial b^\dagger}\bigg)_{b\,=\langle b
\rangle} =\, 0 \,, ~~~~~ b \,=\, \Phi,\eta_1,\eta_2,\zeta \,.  \label{min1}
\end{eqnarray}
For the VEVs of the multiplets, we adopt the notation
\begin{eqnarray}
\langle \zeta\rangle \,=\, \frac{\tilde v}{\sqrt{2}} \,, ~~~~~
\langle \Phi \rangle \,=\, \frac{1}{\sqrt2} \left(\begin{array}{c} 0 \\ v \\
\end{array}\right) , ~~~~~
\langle \eta_1 \rangle \,=\, \frac{1}{\sqrt2} \left(\begin{array}{c} 0 \\ v_1^{} \\
\end{array}\right) , ~~~~~
\langle \eta_2 \rangle \,=\, \frac{1}{\sqrt2} \left(\begin{array}{c} 0 \\ v_2^{} \\
\end{array}\right) ,
\end{eqnarray}
and so in general  $\tilde v$, $v$, $v_1^{}$, and $v_2^{}$ can be zero or nonzero.
We have set the charged components of the doublets to zero in order to preserve
the electromagnetic $U(1)$ symmetry.
To find the minima, we construct the 4$\times$4 Hessian matrix having elements
\,$\partial^2\mathcal{V}/\big(\partial b_m^{}\partial b_n^\dagger\big)$,\,
apply to it the solutions to Eq.\,(\ref{min1}), and require the Hessian to have
a positive determinant and positive eigenvalues.
The minimum with the desired vacuum pattern
\begin{eqnarray} \label{vevs}
\tilde v \,\neq\, 0 \,, ~~~~~~~ v \,\neq\, 0 \,, ~~~~~~~ v_1^{} \,=\, v_2^{} \,=\, 0
\end{eqnarray}
occurs if the parameters in $\cal V$ satisfy the relations in Eq.\,(\ref{mu1mu2}) and
the inequality
\begin{eqnarray} &&
\Big\{ \Big[2\mu_{21}^2 + (\lambda_{31} + \lambda_{41}) v^2 + \lambda_{1\zeta }\tilde v^2 \Big]
\Big[2\mu _ {22}^2 + (\lambda_{32}+\lambda_{42}) v^2 + \lambda _ {2\zeta }\tilde v^2 \Big]
-2 \mu^2_{\eta\zeta}\tilde v^2 \Big\}
\nonumber \\ && \; \times\, \Big\{
\Big[ 2 \lambda_1 v^2+ \lambda_{3\zeta}\tilde v^2+2 \mu^2_1 \Big]
\Big[ \lambda_{3\zeta} v^2+ 2 \lambda_{\zeta}\tilde v^2+2 \mu^2_{\zeta} \Big]
- \lambda^2_{3\zeta} v^2 \tilde v^2 \Big\}
\,>\, 0 \,.
\end{eqnarray}
However, these conditions do not yet guarantee that other minima, with only $v_1^{}$
or $v_2^{}$ being zero or with none of the VEVs being zero, are not lower.
The corresponding expressions in these other cases are lengthy and hence not shown here.
Therefore, to make sure that Eq.\,(\ref{vevs}) corresponds to the absolute minimum of
$\cal V$, in numerical simulations we check that the parameter values yield
the lowest $\cal V$ among the different minima, as well as meet all other requirements.

\section{Masses and interactions of new fermions\label{Nk}}

From ${\cal L}_N$ in Eq.\,(\ref{LN}), we can express the terms responsible for the new
fermions' masses as
\begin{eqnarray}
{\cal L}_N^{} \,\supset\, {\cal L}_N' \,=\, -\tfrac{1}{2} \Big(
\overline{(N_R)\raisebox{1pt}{$^{\rm c}$}} ~~~~~ \overline{N_L} \Big)
\left(\!\begin{array}{cc} \sqrt2\,\tilde v\,\hat{\texttt Y}_1^{} & M_N^{\rm T} \vspace{2pt} \\
M_N^{} & \sqrt2\,\tilde v\,\hat{\texttt Y}_2^{} \end{array}\!\right) \left(\!\begin{array}{c}
N_R^{} \vspace{3pt} \\ (N_L)\raisebox{1pt}{$^{\rm c}$} \end{array}\!\right)
+\; {\rm H.c.}
\end{eqnarray}
This implies that in the presence of $\hat{\texttt Y}_{1,2}$ the left- and right-handed
components of $N_k$ mix, leading to Majorana mass eigenstates ${\texttt N}_{k,L}$ and
${\mathtt N}_{k,R}$ which in general have different masses.
In terms of the latter,
\begin{eqnarray}
{\cal L}_N' &=& -\tfrac{1}{2} \Big[
\overline{{\mathtt N}_L^{}}\,\hat{m}_L^{}\,({\mathtt N}_L^{})\raisebox{1pt}{$^{\rm c}$}
+ \overline{({\mathtt N}_R^{})\raisebox{1pt}{$^{\rm c}$}}\,\hat{m}_R^{}\,{\mathtt N}_R^{} \Big]
\,+\, {\rm H.c.} \,, ~~~~ ~~~
\left(\!\begin{array}{c} N_R^{} \vspace{2pt} \\ (N_L)\raisebox{1pt}{$^{\rm c}$}
\end{array}\!\right) =\,\mathbb{U} \left(\!\begin{array}{c} {\mathtt N}_R^{} \vspace{2pt} \\
({\mathtt N}_L)\raisebox{1pt}{$^{\rm c}$} \end{array}\!\right) ,
\nonumber \\
\mathbb{U} &=& \left(\!\begin{array}{cc} U_{RR}^{} & U_{RL}^{} \vspace{1pt} \\
U_{LR}^{} & U_{LL}^{} \end{array}\!\right) , ~~~~ ~~~
\left(\!\begin{array}{cc} \hat m_R^{} & 0 \vspace{1pt} \\ 0 & \hat m_L^{}
\end{array}\!\right)
=\, {\mathbb U}^{\rm T} \left(\!\begin{array}{cc} \sqrt2\,\tilde v\,\hat{\texttt Y}_1^{} &
M_N^{\rm T} \vspace{1pt} \\ M_N^{} & \sqrt2\,\tilde v\,\hat{\texttt Y}_2^{}
\end{array}\!\right) {\mathbb U} \,,
\end{eqnarray}
where $\mathbb{U}$ is a unitary 6$\times$6 matrix, $U_{LL,LR,RL,RR}$ denote its 3$\times$3
submatrices, and $\hat m_{L,R}^{}$ are each diagonal 3$\times$3 matrices for the eigenmasses.
Hence \,$U_{LL,RR}\to\openone$\, and \,$U_{LR,RL}\to0$\, if $\hat{\texttt Y}_{1,2}$ are
negligible or vanishing.
It follows that the interactions of these fermions with the scalars are described by
\begin{eqnarray} \label{LY}
{\cal L}_N^{} &\supset&
\overline{\ell_{\,}} \big[ \big(c_H^{}{\cal Y}_1^{}U_{RR}^{}-s_H^{}{\cal Y}_2^{}U_{LR}^{}\big) H_1^-
+ \big( s_H^{}{\cal Y}_1^{}U_{RR}^{}+c_H^{}{\cal Y}_2^{}U_{LR}^{}\big) H_2^- \big] {\mathtt N}_R^{}
\nonumber \\ && \! +\;
\overline{\ell_{\,}} \big[ \big(c_H^{}{\cal Y}_1^{}U_{RL}^{}-s_H^{}{\cal Y}_{2}^{}U_{LL}^{}\big) H_1^-
+ \big( s_H^{}{\cal Y}_1^{}U_{RL}^{}+c_H^{}{\cal Y}_2^{}U_{LL}^{}\big) H_2^- \big]
({\mathtt N}_L)^{\rm c}
\nonumber \\ && \! -\; \tfrac{1}{\sqrt2}\, \overline{\nu} \big[
\big(c_S^{}{\cal Y}_1^{}U_{RR}^{}-s_S^{}{\cal Y}_2^{}U_{LR}^{}\big) {\cal S}_1^{} + \big(
s_S^{}{\cal Y}_1^{}U_{RR}^{}+c_S^{}{\cal Y}_2^{}U_{LR}^{}\big) {\cal S}_2^{} \big] {\mathtt N}_R^{}
\nonumber \\ && \! -\; \tfrac{1}{\sqrt2}\, \overline{\nu} \big[
\big(c_S^{}{\cal Y}_1^{}U_{RL}^{}-s_S^{}{\cal Y}_2^{}U_{LL}^{}\big) {\cal S}_1^{} + \big( s_S^{}
{\cal Y}_1^{}U_{RL}^{}+c_S^{}{\cal Y}_2^{}U_{LL}^{}\big) {\cal S}_2^{} \big] ({\mathtt N}_L)^{\rm c}
\nonumber \\ && \! +\; \tfrac{i}{\sqrt2}\, \overline{\nu} \big[
\big(c_P^{}{\cal Y}_1^{}U_{RR}^{}+s_P^{}{\cal Y}_2^{}U_{LR}^{}\big) {\cal P}_1^{} + \big(
s_P^{}{\cal Y}_1^{}U_{RR}^{}-c_P^{}{\cal Y}_2^{}U_{LR}^{}\big) {\cal P}_2^{} \big] {\mathtt N}_R^{}
\nonumber \\ && \! +\; \tfrac{i}{\sqrt2}\, \overline{\nu} \big[
\big(c_P^{}{\cal Y}_1^{}U_{RL}^{}-s_P^{}{\cal Y}_2^{}U_{LL}^{}\big) {\cal P}_1^{} + \big( s_P^{}
{\cal Y}_1^{}U_{RL}^{}+c_P^{}{\cal Y}_2^{}U_{LL}^{}\big) {\cal P}_2^{} \big] ({\mathtt N}_L)^{\rm c}
\nonumber \\ && \! +\;
\frac{s_\xi^{}h-c_\xi^{}\tilde s}{\footnotesize\mbox{$\sqrt2$}} \!\! \begin{array}[t]{l} \Big[
\overline{({\mathtt N}_R)\raisebox{1pt}{$^{\rm c}$}} \big( U_{RR}^{\rm T}\hat{\texttt Y}_1^{}U_{RR}^{}
+ U_{LR}^{\rm T}\hat{\texttt Y}_2^{} U_{LR}^{} \big) {\mathtt N}_R^{}
- 2\,\overline{{\mathtt N}_L^{}} \big( U_{RL}^{\rm T} \hat{\texttt Y}_1^{}U_{RR}^{} +
U_{LL}^{\rm T}\hat{\texttt Y}_2^{}U_{LR}^{} \big){\mathtt N}_R^{}
\\ \;+~
\overline{{\mathtt N}_L^{}} \big( U_{RL}^{\rm T} \hat{\texttt Y}_1^{}U_{RL}^{} +
U_{LL}^{\rm T} \hat{\texttt Y}_2^{}U_{LL}^{} \big) ({\mathtt N}_L)\raisebox{1pt}{$^{\rm c}$} \Big]
\end{array}
\nonumber \\ && \! +\; {\rm H.c.}
\end{eqnarray}

Among the important contributions to the DM annihilation rate are those with lepton pairs,
$\ell\bar\ell'$ or $\nu\nu'$, in the final state which are dominated by tree-level
contributions from diagrams mediated by the inert scalars in the $t$ and $u$ channels.
If the effects of $\hat{\texttt Y}_{1,2}$ can be neglected, we take $N_1$ to be the DM.
The cross section of \,$N_1 N_1\to\ell_o^-\ell_r^+$\, is then
\begin{eqnarray}
\sigma_{N_1^{}N_1^{}\to\ell_o^{}\bar\ell_r^{}}^{} &=&
\frac{1}{32\pi \big(s-4M_1^2\big)} \Bigg\{ \Bigg[
\frac{c_H^2s_H^2{\tt L}_1^{}}{{\cal M}_1^2+\tfrac{1}{2}s}
+ \frac{c_H^2s_H^2 {\tt L}_2^{}}{{\cal M}_2^2+\tfrac{1}{2}s}
+ \frac{\big(c_H^4+s_H^4\big)({\tt L}_1+{\tt L}_2)}{{\cal M}_1^2+{\cal M}_2^2+s}
\Bigg] M_1^2\, {\rm Re}\big({\cal Y}_t^{2*}Y_u^2\big)
\nonumber \\ && \hspace{7em} +\;
c_H^2s_H^2\big(|{\cal Y}_t^{}|^4+|{\cal Y}_u^{}|^4\big) \Bigg( \sqrt{1-\frac{4M_1^2}{s}} +
\frac{{\cal M}_1^4\,{\tt L}_1^{}-{\cal M}_2^4\,{\tt L}_2^{}}{m_{H_1}^2s-m_{H_2}^2s} \Bigg)
\nonumber \\ && \hspace{7em} +\;
\big(c_H^4|{\cal Y}_t^{}|^4+s_H^4|{\cal Y}_u^{}|^4\big){\mathbb R}_1^{}
+ \big(s_H^4|{\cal Y}_t^{}|^4+c_H^4|{\cal Y}_u^{}|^4\big){\mathbb R}_2^{} \Bigg\} \,,
\end{eqnarray}
where
\begin{eqnarray}
{\tt L}_a^{} &=& \ln\frac{2{\cal M}_a^2+s-\sqrt{s^2-4M_1^2s}}
{2{\cal M}_a^2+s+\sqrt{s^2-4M_1^2s}} \;, ~~~ ~~~~
{\cal Y}_t^2 \,=\, ({\cal Y}_1^{})_{o1}^{}({\cal Y}_1^{})_{r1}^* \,, ~~~ ~~~~
{\cal Y}_u^2 \,=\, ({\cal Y}_2^{})_{o1}^{}({\cal Y}_2^{})_{r1}^* \,
\nonumber \\
{\cal M}_a^2 &=& m_{H_a}^2-M_1^2 \,, ~~~~ ~~~
{\mathbb R}_a^{} \,=\, \frac{{\cal M}_a^4+\tfrac{1}{2}\,m_{H_a}^2s}{{\cal M}_a^4+m_{H_a}^2s}\,
\sqrt{1-\frac{4M_1^2}{s}}
+ \frac{{\cal M}_a^2}{s}\,{\tt L}_a^{} \,.
\end{eqnarray}
For \,$o=r$,\, there are also contributions from $h$- and $\tilde s$-mediated diagrams in
the $s$ channel, but these are suppressed by $m_{\ell_o}/v$ and hence can be ignored.
The cross section of \,$N_1 N_1\to\nu_o^{}\nu_r^{}$,\, arising from ${\cal S}_{1,2}$- and
${\cal P}_{1,2}$-exchange diagrams, is much lengthier and not displayed here.

Also potentially important are the channels
\,$N_1^{}N_1^{}\to(h^*,\tilde s^*)\to X_{\mathrm{SM}}=b\bar b,W^+W^-,ZZ,t\bar t$,\,
each of which has a cross section
\begin{eqnarray} \label{N1N1->h,s->SM}
\sigma_{N_1^{}N_1^{}\to(h^*,\tilde s^*)\to X_{\mathrm{SM}}}^{} &=& \frac{c_\xi^2 s_\xi^2}{2}
\Big\{ \big(s-2M_1^2\big) \Big[ \big|\big(\hat{\texttt Y}_1\big){}_{11}^{}\big|^2
+ \big|\big(\hat{\texttt Y}_2\big){}_{11}^{}\big|^2 \Big] + 4 M_1^2\,{\rm Re}\Big[
\big(\hat{\texttt Y}_1\big)_{11}\big(\hat{\texttt Y}_2\big)_{11}^{} \Big] \Big\}
\nonumber \\ && \times\;
\bigg| \frac{1}{s-m_h^2+i\Gamma_h^{}m_h^{}} -
\frac{1}{s-m_{\tilde s}^2+i\Gamma_{\tilde s}^{}m_{\tilde s}^{}} \bigg|^2 ~
\frac{\Gamma_{h_{\mathrm{SM}}\to X_{\mathrm{SM}}}\big|_{m_h^2=s}}{\sqrt{s-4M_1^2}} \,.
\end{eqnarray}
We have also looked at \,$N_1^{}N_1^{}\to hh$,\, but its contribution turns out to be
unimportant in our parameter space of interest.

The scattering of $N_1$ off a nucleon $\cal N$ at tree level is mediated by $h$ and $\tilde s$.
The cross section in the nonrelativistic limit is
\begin{eqnarray} \label{dm-nucleon}
\sigma_{\rm DM\mbox{-}nucleon}^{} \;=\;
\frac{c_\xi^2 s_\xi^2\, g_{{\cal NN}h\,}^2 M_1^2 m_{\cal N}^2\, \Big|
\big(\hat{\texttt Y}_1^{}\big)_{11} + \big(\hat{\texttt Y}_2^*\big)_{11} \Big|^2}
{2\pi\,\big(M_1^{}+m_{\cal N}^{}\big)\raisebox{1pt}{$^2$}}
\bigg( \frac{1}{m_h^2} - \frac{1}{m_{\tilde s}^2} \bigg)^{\!2} \,,
\end{eqnarray}
where the effective Higgs-nucleon coupling \,$g_{{\cal NN}h}^{}=0.0011$\,
is at the lower end of its range estimated in Ref.\,\cite{He:2011gc} and thus helps
minimize the prediction in light of the strict experimental limits.

\end{document}